\documentclass[twocolumn,twocolappendix]{aastex631} %linenumbers
\usepackage{amsmath,amssymb,amsfonts,bm}
\usepackage{epstopdf}
\usepackage{epsfig}
\usepackage{natbib,caption2}
\usepackage{graphicx}   % Including figure files
\usepackage{float}
\usepackage{mhchem}
\usepackage{longtable}
\usepackage{graphics}
\usepackage{hyperref}
\usepackage{color}
\usepackage{calc}
\usepackage{threeparttable}

\newcommand \beq{\begin{equation}}
\newcommand \eeq{\end{equation}}
\newcommand \bey{\begin{eqnarray}}
\newcommand \eey{\end{eqnarray}}

\usepackage{natbib}

\newcommand{\gsim}{\lower.5ex\hbox{$\; \buildrel > \over \sim \;$}}
\newcommand{\lsim}{\lower.5ex\hbox{$\; \buildrel < \over \sim \;$}}

\shortauthors{Bian et al.}
\begin{document}

\title{Two Channels of Metal-Rich Compact Stellar System Formation: Starbursts under High Ram Pressure versus Tidal Stripping}

\correspondingauthor{Min Du}
\email{dumin@xmu.edu.cn}

\author{Yuan Bian}
\affiliation{Department of Astronomy, Xiamen University, Xiamen, Fujian 361005, People’s Republic of China}

\author{Min Du}
\affiliation{Department of Astronomy, Xiamen University, Xiamen, Fujian 361005, People’s Republic of China}

\author{Victor P. Debattista}
\affiliation{Jeremiah Horrocks Institute, University of Central Lancashire, Preston PR1 2HE, UK}

\author{Dylan Nelson}
\affiliation{Universit\"{a}t Heidelberg, Zentrum f\"{u}r Astronomie, ITA, Albert-Ueberle-Str. 2, 69120 Heidelberg, Germany}

\author{Mark A. Norris}
\affiliation{Jeremiah Horrocks Institute, University of Central Lancashire, Preston PR1 2HE, UK}

\author{Luis C. Ho}
\affiliation{Kavli Institute for Astronomy and Astrophysics, Peking University, Beijing 100871, People’s Republic of China}
\affiliation{Department of Astronomy, School of Physics, Peking University, Beijing 100871, People’s Republic of China}

\author{Shuai Lu}
\affiliation{Department of Astronomy, Xiamen University, Xiamen, Fujian 361005, People’s Republic of China}

\author{Renyue Cen}
\affiliation{Center for Cosmology and Computational Astrophysics, Institute for Advanced Study in Physics, Zhejiang University, Hangzhou 310027, People’s Republic of China}
\affiliation{Institute of Astronomy, School of Physics, Zhejiang University, Hangzhou 310027, People’s Republic of China}

\author{Shuo Ma}
\affiliation{Department of Astronomy, Xiamen University, Xiamen, Fujian 361005, People’s Republic of China}

\author{Chong Ge}
\affiliation{Department of Astronomy, Xiamen University, Xiamen, Fujian 361005, People’s Republic of China}

\author{Taotao Fang}
\affiliation{Department of Astronomy, Xiamen University, Xiamen, Fujian 361005, People’s Republic of China}

\author{Hui Li}
\affiliation{Department of Astronomy, Tsinghua University, Haidian DS, Beijing 100084, People’s Republic of China}

\begin{abstract}

Most galaxies follow a well-defined scaling relation between metallicity and stellar mass; however, some outliers at the low-mass end of the observed galaxy population exhibit unusually high metallicity for their mass. Understanding how these objects get to be so metal-rich is vital for understanding the role of feedback in galaxy formation. Using the TNG50 simulation, we explore the origins of this phenomenon. We identify 227 metal-rich, compact stellar systems (CSSs) that deviate significantly from this scaling relation. These CSSs are satellites located in the vicinity of massive host galaxies, with stellar masses ranging from $10^{8} M_{\odot}$ to $10^{10} M_{\odot}$ (including six systems that are close analogs of the M31–M32 system). Contrary to the previously assumed scenario that such objects are predominantly products of tidal stripping, we find that more often ram pressure plays a major role in their formation. Indeed, 76\% (173) of these CSSs are formed through a burst of star formation occurring around the time of the first pericentric passage, typically at redshifts $z\lesssim1$, aided by strong ram pressure and tidal forces. The high ram pressure, resulting from the CSSs’ rapid motion near the host halo center, facilitates metal enrichment, producing high-metallicity CSSs by confining the metal-rich gas from bursty star formation, which leads to distinct stellar populations characterized by enhanced metallicity and high $\alpha$-abundance. The other 24\% (54) of metal-rich CSSs are generated through the tidal stripping of massive progenitors. Our results further indicate that M32 is more likely to have formed through intense star formation events rather than through gradual tidal stripping.

\end{abstract}


% \linenumbers

\section{Introduction}

Over the past three decades, increasing numbers of compact stellar systems (CSSs) have been discovered. These intriguing objects occupy the parameter space between classical globular clusters (GCs) and normal galaxies. Based on previous studies, both ultracompact dwarf galaxies \citep[UCDs;][]{Drinkwater2004,Hasegan2005,Jones2006,Wehner&Harris2007,Misgeld2008,Mieske2009,Madrid2010,Brodie2011} and compact elliptical galaxies \citep[cEs;][]{Price2009,Chilingarian2009,Chilingarian2010,Norris2014,Janz2016,Kim2020} fall under the broader category of CSSs, with UCDs extending above GC masses ($M_{*} \gtrsim 10^{6}$–$10^{8} M_{\odot}$) and cEs those objects at still higher masses ($\gtrsim 10^{9} M_{\odot}$). Notably, some objects at the higher mass range ($M_{*} = 10^{8}$–$10^{10} M_{\odot}$) exhibit enhanced stellar metallicities, typically deviating by about 0.4–0.6 dex from the stellar mass–metallicity relation observed in ordinary dwarf galaxies of comparable mass \citep{Gallazzi2005} and more typical of galaxies 10 times more massive.

A commonly invoked formation scenario for metal-rich CSSs is the heavy stripping of initially more massive galaxies \citep{Bekki2001,Brodie2011,Janz2016,Kim2020,Gallazzi2021,Mayes2021,Deeley2023}. CSSs caught in the act of formation and still embedded in tidal streams of stars from their disrupted progenitors provide some direct evidence supporting this scenario \citep{Chilingarian2009,Huxor2011,Chilingarian2015,Wang2023}. In this process, the least bound stars in the outer parts, which have lower metallicity, are preferentially stripped, leaving behind a metal-rich compact remnant. The resulting objects deviate from the mass–metallicity relation \citep[see also][]{Chilingarian2009} by virtue of the high metallicity of the bulge of their more massive progenitors. However, the extremely high metallicities of outliers in the mass–metallicity relation are only typical of the innermost regions ($\sim r_{\mathrm{e}}/8$) of massive galaxies \citep{McDermid2015}. Thus, for instance, a Milky Way–sized progenitor would be necessary to generate M32 analogs if only tidal stripping (TS) were responsible for their formation. Yet M31’s stellar halo is much less massive \citep[$\sim 10^{9}M_{\odot}$ level; see][]{Ibata2007,Courteau2011,Williams2012,Gilbert2012}, and no clear evidence of such a massive progenitor for M32 exists within M31 \citep{DSouza2018}.

An alternative hypothesis posits the direct formation of metal-rich CSSs through starbursts triggered within their host galaxy environments. Interactions with host galaxies can induce gas inflows toward the centers of gas-rich dwarfs \citep{Graham2002}. \citet{Du2019} showed that the high ram pressure experienced during rapid passages near massive host galaxies promotes star formation through gas compression, increasing the stellar density in the central region, while also confining metals within the dwarf, leading to the rapid enrichment of new stellar populations. Independently, \citet{Williamson2018} obtained similar results through wind tunnel experiments, also finding that ram pressure confines gas and metals to dwarf galaxies, which would otherwise have escaped their shallow potentials. In addition, cosmological zoom-in simulations, such as the FIRE simulations, which resolve scales below individual starforming clouds, support this picture \citep[e.g.,][]{Sparre2017, Feldmann2017, Ma2018}. These studies show that bursty star formation (BSF), typical in dwarfs and high-redshift galaxies, is followed by energetic feedbackdriven outflows and gas infall on timescales shorter than 100 Myr. The BSF features have also been noticed in many other studies \citep[see][and references therein]{Sales2022}, though the underlying physics of these features remains poorly understood. For this work, we term this scenario the BSF scenario to differentiate it from the TS scenario to explain the formation of CSSs—characterized by rapid, short-lived bursts of star formation, primarily driven by environmental factors in the vicinity of a more massive host. This process compresses gas and induces bursts of star formation over timescales typically spanning 100–200 Myr, leading to substantial variability in star formation rates and often showing a strong peak.

While many studies have addressed the formation and evolution of low-mass dwarf satellites \citep[see][and references therein]{Mayer2010,deAlmeida2024}, the formation of the metal-rich CSSs (i.e., outliers in the mass-metallicity relation in the $10^{8}-10^{10}\ M_{\odot}$ mass range) has not been studied in a cosmological context. In this work, we quantitatively assess the relative likelihood of different formation pathways of metal-rich CSSs using a large-volume cosmological simulation.

The outline of this Letter is as follows: We introduce the IllustrisTNG simulation and sample selection in Section~\ref{sec:data}. In Section~\ref{MZR Evolution}, we present CSSs selected based on our criteria in TNG50, along with their mass–metallicity evolutionary trajectories. Section~\ref{Evolution} shows the star formation history and chemical enrichment history. In Section~\ref{Orbit}, we present the orbital characteristics of our CSSs. In Section~\ref{CSS-BSF}, we provide chemodynamics information on the surrounding gas properties of these CSSs. Section~\ref{Fingerprint} gives the information of the features of our CSSs in $[\alpha/\text{Fe}]-\text{[Fe/H]}$ space. We discuss our results in Section~\ref{sec:discussion}. Finally, in Section~\ref{sec:summary}, we summarize our main results and conclusions.

\section{I\lowercase{llustris}TNG Simulations and Sample Selection}
\label{sec:data}

The IllustrisTNG Project \citep[hereafter TNG;][]{Marinacci2018,Naiman2018,Nelson2018,Pillepich2018a,Springel2018} is a suite of magnetohydrodynamic cosmological simulations run with the moving-mesh code AREPO \citep{Springel2010}. The suite encompasses three distinct runs, TNG50, TNG100, and TNG300, characterized by cubic volumes of approximately 50, 100, and 300 Mpc side lengths, respectively. The simulations are run with gravo-magnetohydrodynamics (MHD) and incorporate a comprehensive subgrid model \citep[see][for details]{Weinberger2017,Pillepich2018b}. TNG has successfully reproduced many fundamental properties and scaling relations of observed galaxies. In particular, \citet{Nelson2019} highlight TNG50’s ability to realistically reproduce galactic outflows due to supernovae and black hole feedback. TNG50’s high resolution facilitates detailed studies, with \citet{Pillepich2019} noting the consistency in starforming galaxy thickness compared to observations. Additionally, the simulated barred galaxies \citep{Zhao2020} further testify to the reliability of TNG100. TNG galaxies are identified and characterized with the Friends-of-Friends \citep[\texttt{FoF}][]{Davis1985} and \texttt{SUBFIND} \citep{Springel2001} algorithms. Halos are dark matter regions containing galaxies and consist of a smooth component and gravitationally bound subhalos. Merger trees for subhalos are constructed using the SubLink algorithm, which identifies descendants through a three-step process based on shared particles and a merit function; the algorithm also addresses undetected subhalos by allowing snapshot skips and determines the main progenitor based on its mass history \citep[see][for details]{Rodriguez-Gomez2015}. Resolution elements (gas, stars, dark matter, and black holes) belonging to an individual galaxy are gravitationally bound to its host subhalo. The adopted cosmology of the TNG project is the Planck 2015 \citep{Planck2016}, given by: $\Omega_{\mathrm{\Lambda}} = 0.6911$, $\Omega_{\mathrm{m}} = 0.3089$, $\Omega_{\mathrm{b}} = 0.0486$, $\sigma_{8} = 0.8159$, $n_{s} = 0.9667$, and $h = 0.6774$.

We use the TNG50 \citep[aka TNG50-1;][]{Nelson2019,Pillepich2019} simulation, for its exceptional resolution and statistical significance, making it highly suitable for studying the evolution of dwarf galaxies. TNG50 has $2 \times 2160^3$ initial resolution elements (baryon mass of $8 \times 10^4\ M_{\odot}$). Dark matter particles are resolved at masses of $4.5 \times 10^5 M_{\odot}$. The size resolution is defined by the softening lengths of the collisionless components (dark matter and stars), which scale with cosmic time, starting at 576 comoving pc until $z = 1$, after which they are fixed at a physical size of 288 pc \citep{Pillepich2019}. The gas softening length begins at 74 comoving pc, with the smallest gas cell size being 8 pc. This resolution approaches or exceeds that of modern ``zoom-in" simulations of individual massive galaxies, while the volume  contains $\sim20,000$ resolved galaxies with $M_{*} > 10^{7}\ M_{\odot}$.

\subsection{Sample Selection}\label{Sample Selection}

\begin{figure*}
    \centering
    \includegraphics[width=1\linewidth]{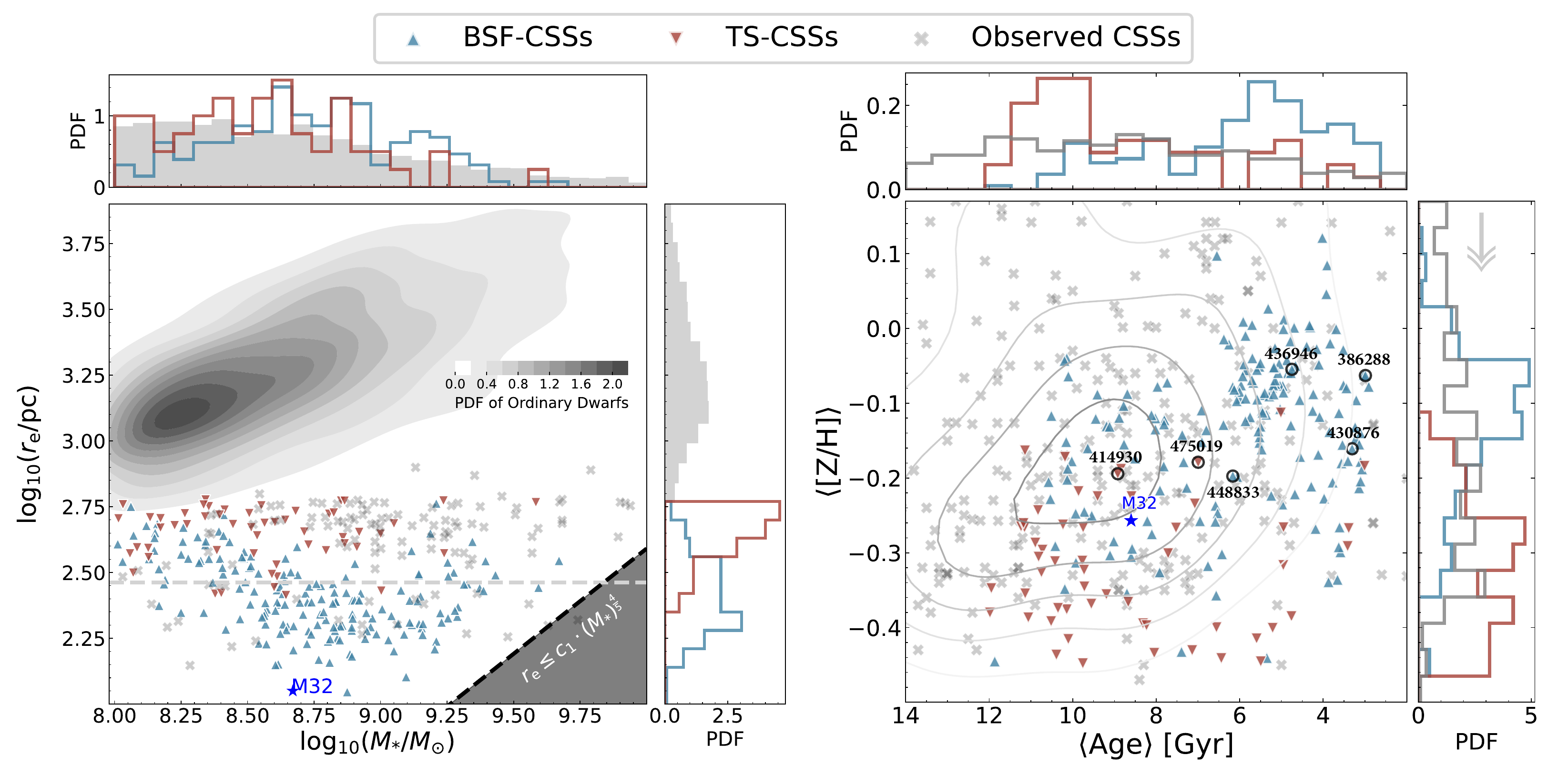}
    \caption{Left: the mass–size distribution of CSSs in TNG50 at $z = 0$. The blue and red triangles represent the CSSs in TNG50 selected based on our criteria. Blue triangles denote BSF-CSSs, and red triangles represent TS-CSSs. The distribution of ordinary dwarf galaxies is indicated by the gray density scale since they are numerous. Observed CSSs \citep{Gurou2015,Chilingarian2015,Janz2016,Ferr2018,Kim2020} are overlaid using gray crosses. The classifications are described in the text following Figure~\ref{classification}. M32 is represented by a blue star. The black dashed line corresponds to the  minimum size limit of stellar systems, and the dark-gray filled region represents the zone of avoidance \citep[\( r_{e} \leq 2.24 \times 10^{-6} \mathrm{pc} \cdot \left(\frac{{M_{{*}}}}{{M_{{\odot}}}}\right)^{4/5} \); see][for details]{Burstein1997,Misgeld2011b}. The horizontal dashed line corresponds to a Plummer-equivalent gravitational softening length at $z = 0$. In the upper and right subpanels, $M_{*}$ and $r_{e}$ distributions are shown as histograms, as well as that of ordinary dwarf galaxies (gray). Right: mass-weighted stellar age vs. metallicity of  TNG50 CSSs. The M32-like samples are marked by black circles. Gray contours show the overall distribution of CSSs from observations. The age and shifted metallicity distributions are shown as histograms in the upper and right panels. The gray arrow in the upper corner of the right subpanel indicates that we apply a uniform 0.75 dex rescaling to the metallicities of TNG50 galaxies to align them with the observed scaling relation \citep{Kirby2013}, as in Figure~\ref{classification}.}
    \label{mass-size}
\end{figure*}

From TNG50, which at $z = 0$ samples thousands of 1081010 Me galaxies, we select the metal-rich CSSs based on their $z = 0$ properties. Our selection criteria do not impose any conditions on their evolution.

We use the selection criteria presented and motivated in \citet{Norris2014} and already used by \citet{Kim2020} and \citet{Chen2022}. To be selected as a metal-rich CSS, a TNG50 galaxy must meet the following three conditions at $z=0$:
\begin{enumerate}
    \item[(A)] \textit{Compactness}: Effective radius $r_{e} \lesssim 600\ \mathrm{pc}$.

    \item[(B)] \textit{Galaxy stellar mass}: $M_{*} = 10^{8}-10^{10}\  M_{\odot}$, calculated within $r_{e}$.

    \item[(C)] \textit{Metallicity}: Adjusted $\langle\text{[Z/H]}\rangle > -0.45$ (Note: The adjustment of $\langle\text{[Z/H]}\rangle$ by -0.75 dex to align with observational data is detailed in Section~\ref{Rescaling of Metallicity}). This additional condition is aimed at identifying CSSs that deviate from the mass-metallicity relation of TNG50 galaxies.
\end{enumerate}
We further exclude objects younger than 3 Gyr because their very limited number of old star particles makes it difficult to trace their evolution using only star particles at $z=0$. The above selection procedures result in a sample of 227 CSSs in the TNG50 simulation. The selected CSSs (triangles in Figure~\ref{mass-size}) are outliers from the mass–size relation of ordinary dwarf galaxies (gray contours). These ordinary dwarfs consist of galaxies other than CSSs in the same mass range. BSF-CSSs (blue triangles) are generally more metal-rich and younger than TS-CSSs (red triangles). TS-CSSs have metal-rich but older stellar populations, similar to those typically found in highmass early-type galaxies \citep{Chilingarian2009}, betraying their genesis within the centers of massive galaxies. These two types of CSSs are similar in mass, while BSF-CSSs are generally more compact than TS-CSSs (left panel of Figure~\ref{mass-size}).

\subsubsection{Rescaling of Metallicity in TNG50} \label{Rescaling of Metallicity}
We adopt a solar metallicity value of $Z_{\odot} = 0.0127$, with solar abundance values for other elements taken from \citet{Asplund2009}. By fitting the mass–metallicity relation for  ordinary dwarf galaxies with $M_{*} = 10^{8}-10^{10}\  M_{\odot}$, we derive $\langle[\mathrm{{Z/H}}]\rangle = 0.17 \times \log_{{10}}\left(\frac{{M_{{*}}}}{{M_{{\odot}}}}\right) - 1.46$, which is overall higher than the universal stellar mass–stellar metallicity relation for dwarf galaxies as reported by \citet{Kirby2013}. To align the metallicities of ordinary dwarf galaxies in TNG50 with the observational data from \citet{Kirby2013} for a similar mass range, we apply a uniform downward adjustment of -0.75 dex to the metallicities of all galaxies within the $M_{*} = 10^{8}-10^{10}\  M_{\odot}$ mass range.

Since detailed comparisons between TNG50 metallicities and the observational data cited in \citet{Kirby2013} have not been made and would need to account for observational biases and differences in calibration methods, this adjustment serves to make our analyses more comparable overall. Importantly, this rescaling adjusts the absolute values but preserves the relative differences, ensuring that the selected CSSs remain more metal-rich than ordinary dwarfs within the same mass range. Furthermore, as shown in the right panel of Figure~\ref{mass-size}, the age–metallicity distribution in TNG50 aligns well with observations post-adjustment, supporting the consistency of our approach.

\subsubsection{M31-M32 Analogs in TNG50}
In our sample of TNG50 CSSs, we identify 131 galaxies  with stellar masses ranging from $10^{8.5}$ to $10^{9.1}\ M_{\odot}$, similar to those of M32. We then get six M32 analogs located in environments closely resembling the M31–M32 system by restricting the stellar mass of the host galaxies to $M^{\mathrm{host}}_{*} = 10^{10.9}-10^{11.2} M_{\odot}$ \citep{Tamm2012,Sick2014}. Their ages and metallicities are highlighted by black circles in the right panel of Figure~\ref{mass-size}. Among the six M32 analogs, ID 414930, 448833, and 475019 also have similar ages and metallicities to M32.

\subsection{Merger Trees of CSSs via Star Particles Matching}\label{Fix}

The standard \texttt{SubLink} algorithm typically relies on dark matter particles to track progenitors. However, this method fails for CSSs owing to their low (or even absent) dark matter content. Instead, we use the unique IDs of star particles in each CSS at $z = 0$ to trace back their progenitors across snapshots. Specifically, we identified progenitors by matching star particle IDs from each CSS with those in earlier snapshots, selecting galaxies that hosted the largest number of these particles. This approach allowed us to construct merger trees for 147 CSSs not tracked by the default method, ensuring the evolutionary continuity of our CSS sample. We implement a matching  threshold of $10^7\ M_{\odot}$, roughly corresponding to 100 star particles, to validate the gravitational association of these particles with their respective galaxies throughout the simulation timeline. Our method effectively identifies progenitors that share significant common star particles with CSSs, offering a robust alternative to traditional dark-matter-based tracking.

\subsection{Identifying CSSs Host Galaxies in TNG50}

To streamline the identification of CSS host galaxies in TNG50 at $z = 0$, we selected galaxies with stellar masses above $10^{9} \ M_{\odot}$ and used spatial coordinates to conduct a proximity search for potential hosts. We locate the 10 nearest neighboring galaxies for each CSS as the potential host. The highest-mass galaxy among these neighbors was initially considered as the potential host. If a CSS spent most of its evolution time outside of this galaxy ($D>R_{\mathrm{vir}}$), the next-highest-mass galaxy was evaluated as a potential host. To enhance the analysis of galaxy trajectories in TNG50, we use cubic spline interpolation to address the simulation's relatively low temporal resolution. This approach generates a detailed time series, distributed across 1000 uniform points, which significantly improves the resolution of the trajectory plots. While the majority of CSSs are hosted by central galaxies in TNG50, there are still 31 CSSs whose host galaxies are identified as satellite galaxies of larger mass.

\section{Results}
\subsection{Two Formation Channels of Metal-Rich CSSs in TNG50}\label{MZR Evolution}

\begin{figure}
    \centering
    \includegraphics[width=1\linewidth]{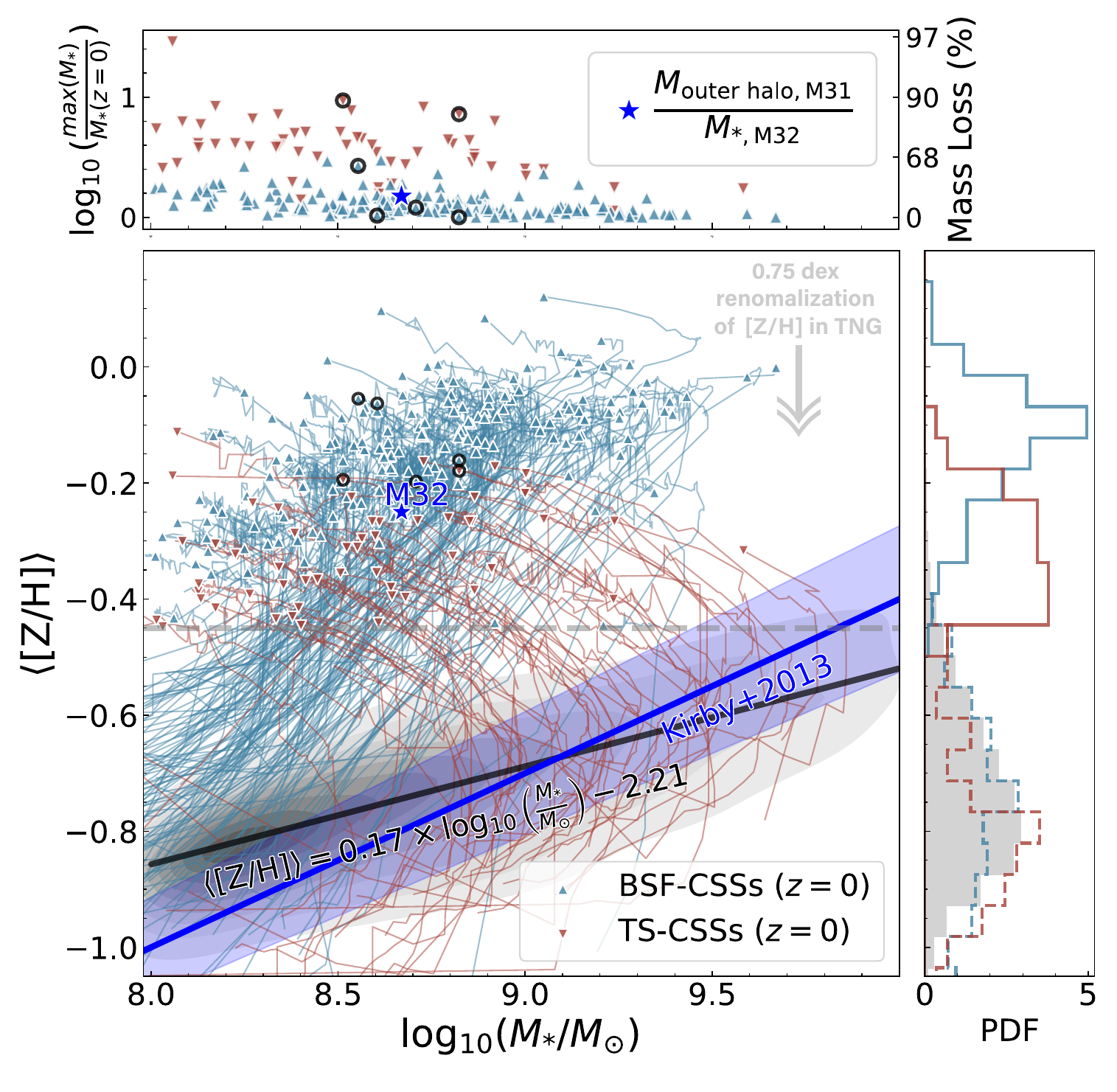}
    \caption{Evolution of CSSs on the stellar mass-metallicity diagram and their classification. Thin trajectories represent individual galaxies that end up in the region of CSSs. The CSSs selected according to our criteria at $z=0$ in TNG50 can be classified as BSF-CSSs (blue triangles) or TS-CSSs (red triangles). The black solid line represents the fitted data for ordinary dwarf galaxies in TNG50. The gray arrow in the upper right corner of the main panel indicates the uniform 0.75 dex rescaling we apply to the metallicities of TNG50 galaxies in this figure, to better align them with the observed scaling relation \citep{Kirby2013}. We calculate the stellar mass of each CSS within the effective radius, $r_{\mathrm{e}}$. For comparison, M32 is represented by the blue star. We approximate M32's mass within $r_{e}$ based on the total mass from \citet{Kormendy&Ho2013}. The selected M32-like CSSs in TNG50 are indicated by black circles. The distribution of ordinary dwarf galaxies at $z=0$ is plotted as the gray-shaded region. The adjusted metallicity distributions at $z=0$ (solid lines) and at the time of the earliest progenitor according to the merger tree (dashed lines), as well as the metallicity distributions of ordinary dwarfs at $z=0$, are shown as histograms in the right panel. In the top panel, we show the ratio of maximum stellar mass to the mass at $z=0$ for all CSSs, with the mass loss fraction shown on the right-hand axis. The blue star in this panel represents the mass ratio between the outer stellar halo of M31 \citep[$M_{\mathrm{outer\ halo,\ M31}}$, e.g.,][]{Gilbert2012} and the stellar mass of M32 ($M_{*,\ \mathrm{M32}}$). Other legends are consistent with the main panel.}
    \label{classification}%
\end{figure}

The 227 CSSs identified in the TNG50 simulation stand out as significant outliers, approximately $2\sigma-3\sigma$ above the typical stellar mass–metallicity relation for ordinary dwarf galaxies, as shown by Figure~\ref{classification} (triangles). We then track the evolution of these CSSs on the stellar mass–metallicity ($M_{*}$-$\langle\text{[Z/H]}\rangle$) plane  over time (individual thin trajectories in Figure~\ref{classification}). We thus categorize them into two distinct types based on whether they follow the trend of ordinary dwarf galaxies on the massmetallicity relation before reaching peak mass on the CSS track: BSF-dominated CSSs (BSF-CSSs; blue triangles; accounting for 173/227; 76\%) that quickly evolve to higher masses and metallicities before experiencing a moderate decrease in mass, and TS-dominated CSSs (TS-CSSs; red triangles; accounting for 54/227; 24\%) that reach high metallicity on the $M_{*}$-$\langle\text{[Z/H]}\rangle$ plane, where metallicity  increases as TS substantially reduces their mass.

\begin{figure*}
    \centering
    \includegraphics[width=1\linewidth]{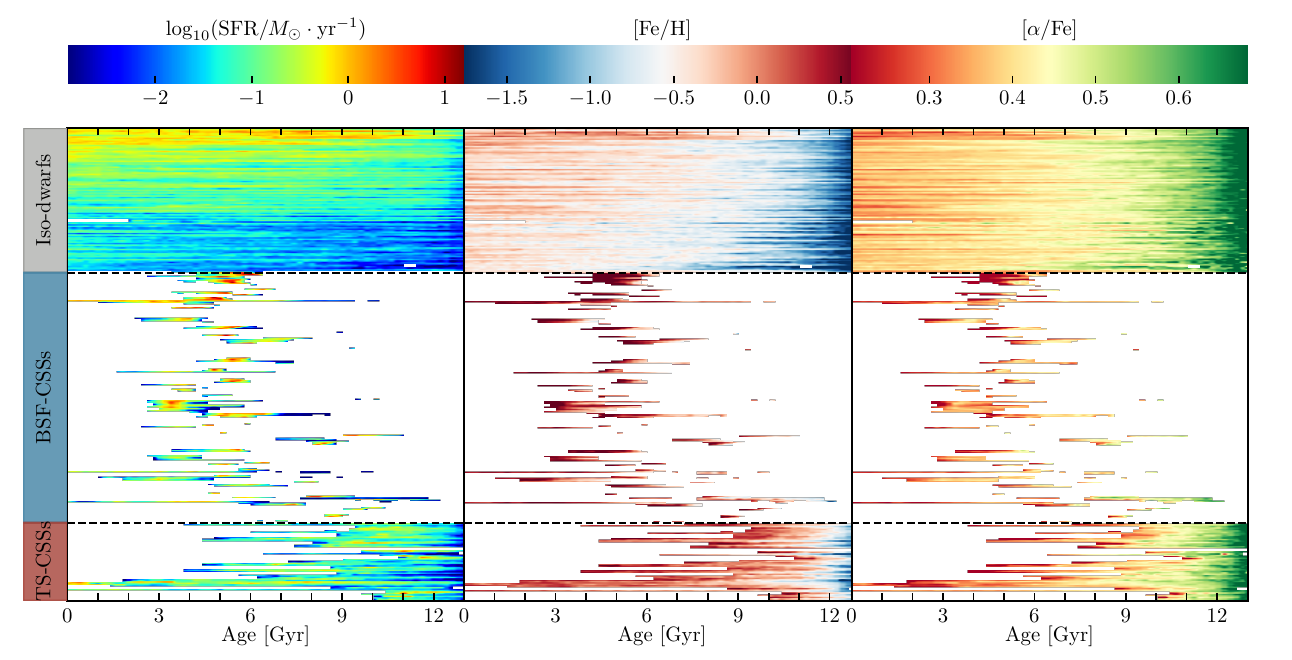}
    \caption{Evolution of each CSS recovered from their stellar population at $z=0$. Each row represents a galaxy, with panels from top to bottom displaying 100 randomly selected Iso-dwarfs (top), BSF-CSSs (middle), and TS-CSSs (bottom). From left to right, the panels show the star formation history, $[\mathrm{Fe}/\mathrm{H}]$, and $[\alpha/\mathrm{Fe}]$.}
    \label{Age_Fe_alpha}%
\end{figure*}

\begin{figure}
    \centering
    \includegraphics[width=1\linewidth]{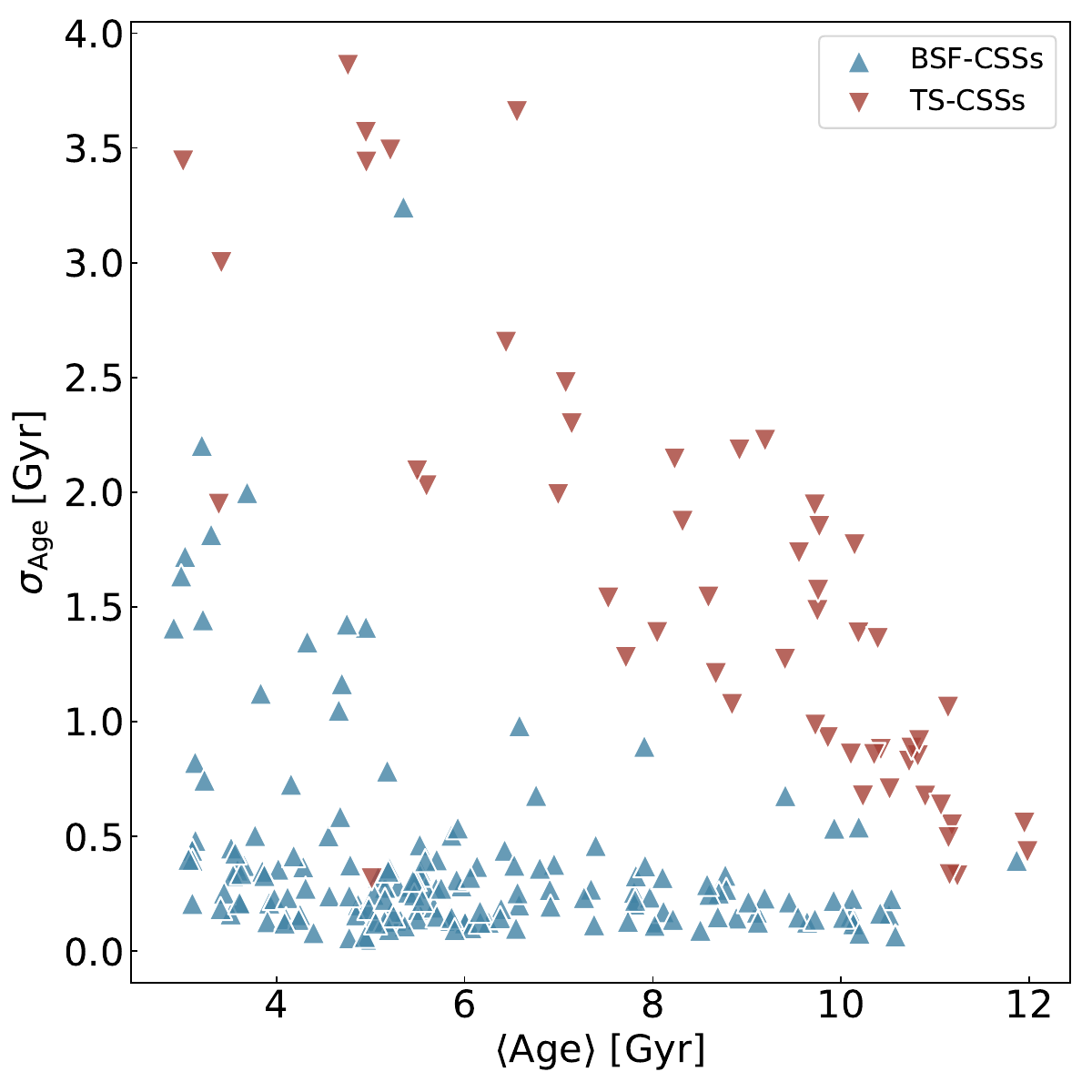}
    \caption{$\langle \text{Age} \rangle$ vs. $\sigma_{\text{Age}}$ of CSSs. We plot the mass-weighted average age, $\langle \text{Age} \rangle$, versus the age dispersion, $\sigma_{\text{Age}}$, for each CSS at $z=0$. Red triangles represent TS-CSSs, while blue triangles represent BSF-CSSs. BSF-CSSs are younger and have a narrower age distribution (typically with $\sigma_{\text{Age}} < 1$ Gyr) compared to TS-CSSs.}
    \label{age_sigmaage}%
\end{figure}

Figure~\ref{Age_Fe_alpha} shows the star formation history along with the enrichment history of $[\mathrm{Fe}/\mathrm{H}]$ and $[\alpha/\mathrm{Fe}]$ for each CSS in our sample, reconstructed from the stellar populations at $z = 0$, compared to a randomly selected group of 100 isolated ordinary dwarf galaxies (Iso-dwarfs). Figure~\ref{age_sigmaage} shows the characteristics of the stellar age distribution for each CSS. All BSF-CSSs experienced rapid star formation bursts (Myr timescale), which are accompanied by high $[\mathrm{Fe}/\mathrm{H}]$ and $[\alpha/\mathrm{Fe}]$ in stars compared to isolated dwarf galaxies. This rapid formation results in BSF-CSSs having a narrower age distribution compared to isolated dwarf galaxies and TS-CSSs, as illustrated in Figure~\ref{age_sigmaage}. In contrast, TS-CSSs exhibit more extended star formation histories with relatively smoother metal enrichment. The distinct signatures of CSSs can potentially be measured using full spectral fitting methods, without the need for data with exceptionally high spatial resolution. This suggests that a detail analysis of integrated spectra could reveal key features of the formation period and subsequent evolution of CSSs.

For these BSF-CSSs, we traced their main progenitors using the star particles at $z = 0$. The progenitors of BSF-CSSs are typically much less massive dwarf galaxies, making it difficult to reliably trace their formation back to even earlier times, due to the limitations of the simulation's resolution. To address this, we set a threshold of approximately $10^7\ M_{\odot}$ (about 100 star particles) for the minimum number of matched particles beyond this limit, only a very small number of particles could be reliably traced.

To gain further insights, we attempted to track BSF-CSSs using not only the star particles at $z = 0$ but also those present in their progenitors across successive snapshots. This method allows us to trace progenitors backward in time, providing additional clues about their earlier evolutionary stages. Our analysis reveals that the progenitors of 34 BSF-CSSs originated as material stripped from satellite galaxies orbiting central galaxies. These stripped fragments continued their orbital trajectories around the central galaxies. Over time, due to the stronger dynamical friction acting on the more massive satellite galaxies, some of these fragments eventually merged with the central galaxies. However, for the majority of BSF-CSS progenitors, the earliest identifiable stage shows them as bound satellites within the halo of the central galaxy. This diversity in origins underscores the complex nature of BSF-CSS progenitors. Importantly, whether they formed as stripped debris or as independent infalling systems, the starburst activity and environmental interactions that drive the formation of BSF-CSSs remain consistent across the population.

\begin{figure*}
    \centering
    \includegraphics[width=1\linewidth]{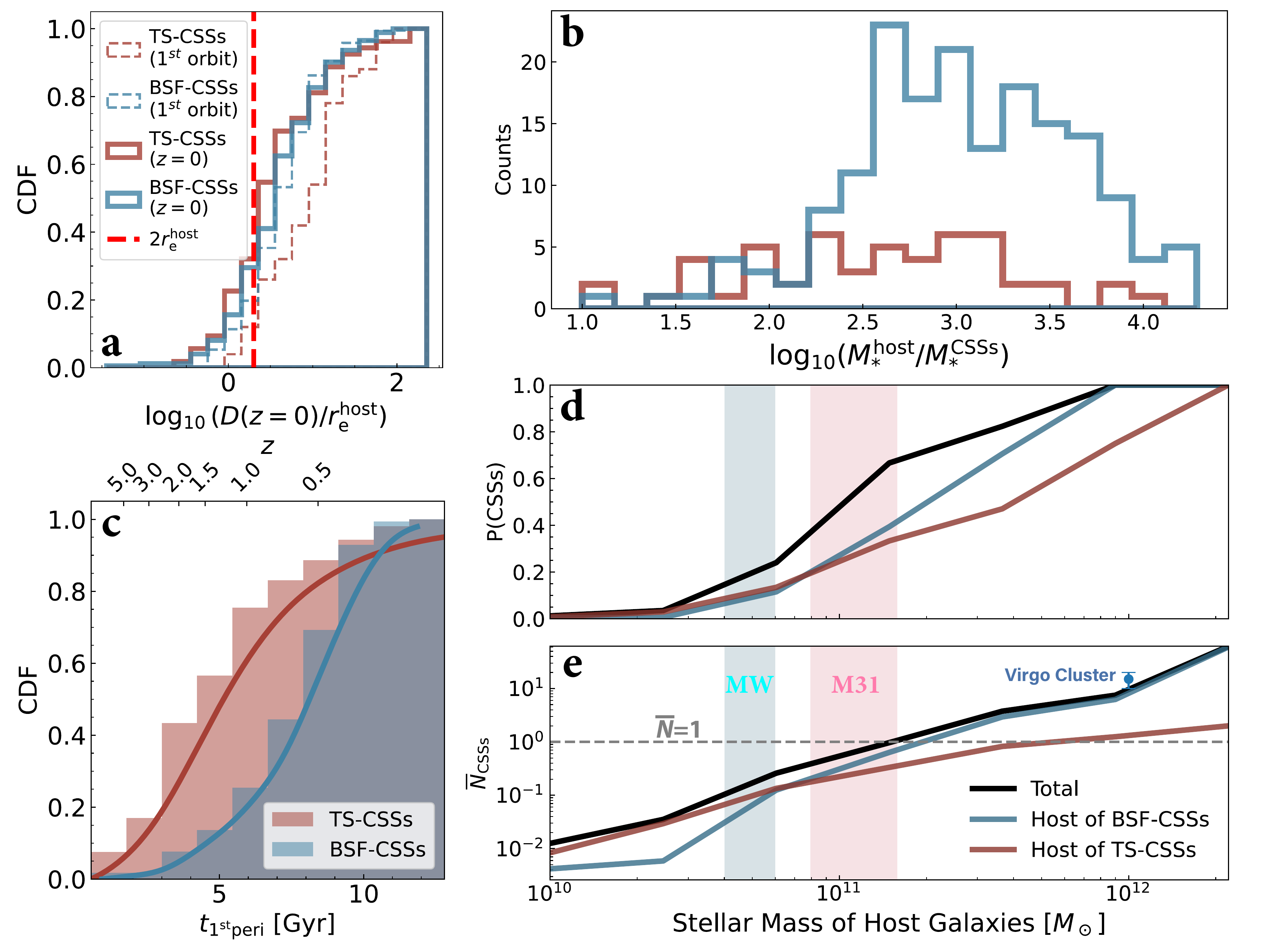}
    \caption{(a) Normalized distribution of the galactocentric distances of CSSs from their hosts at $z=0$. BSF-CSSs are represented in blue, while TS-CSSs are depicted in red. Dashed histograms show the median distance of each CSSs from its host galaxy during the first orbit. (b) The mass ratio of CSSs to their host galaxies' stellar mass at $z=0$ in TNG50. CSSs selected based on our criteria are exclusively found near massive host galaxies. (c) The cumulative distribution function of the time of the first pericentric passage for BSF-CSSs and TS-CSSs. (d) The probability of central galaxies hosting CSSs in TNG50. The black solid line displays the ratio between the number of central galaxies having at least one CSS satellite of $M_{*}=10^{8}-10^{10} M_{\odot}$ and the number of all central galaxies at a given stellar mass. We indicate the stellar mass at $z=0$ of the MW \citep[in cyan, based on][]{Bland-Hawthorn2016} and M31 \citep[pink, from][]{Tamm2012,Sick2014}. (e) The average number of CSSs as a function of the stellar mass of their host galaxies. The number of CSSs hosted by galaxies with a mass around $10^{12} M_{\odot}$ in the Virgo cluster is indicated by the blue circles in this panel.}
    \label{host}
\end{figure*}

We have confirmed that all CSSs selected based on our criteria are satellite galaxies, primarily associated with massive host galaxies. The majority ($\sim 75$\%) of these CSSs are found within $0.1R_{\text{vir}}$ of their host halos but formed outside $2r^{\mathrm{host}}_{e}$ host of  their host galaxies, as illustrated in Figure~\ref{host}(a). Panel (b) shows that all these CSSs have significantly more massive hosts, with  an average halo mass of $\sim 10^{13}\ M_{\odot}$. This is consistent with the observation \citep[e.g.,][]{Norris2014,Janz2016,Ferr2018} that CSSs are often associated with massive galaxies.

In general, BSF-CSSs experience their first pericentric passage later than TS-CSSs, suggesting more rapid transformations (Figure~\ref{host}(c)). In panel (d) we show the probability that any given central galaxy in TNG50 hosts at least one satellite CSS. Approximately half of M31-mass galaxies host at least one CSS, in contrast to around 20\% of Milky Way–mass galaxy hosts. Panel (e) examines how the average number of CSSs varies with the host stellar mass. Over half of the CSSs in our sample are hosted by galaxies with stellar masses exceeding M31. Combining panels (d) and (e), it becomes clear that if TS were the only mechanism involved (represented by the red line), the frequency of CSSs would be too low at the high-mass end; for instance, in the Virgo Cluster, there are 10–20 CSSs hosted by galaxies with $M_{*} \simeq 10^{12}\ M_{\odot}$ \citep{Ferrarese2020}, a number consistent with the CSS predictions (black line) from TNG, while the TS prediction would be for just over one CSS for galaxies with masses larger  than $10^{12} M_{\odot}$.

\subsection{Differences in Evolutionary Pathways}
\label{Evolution}

\begin{figure*}
    \centering
    \includegraphics[width=0.80\linewidth]{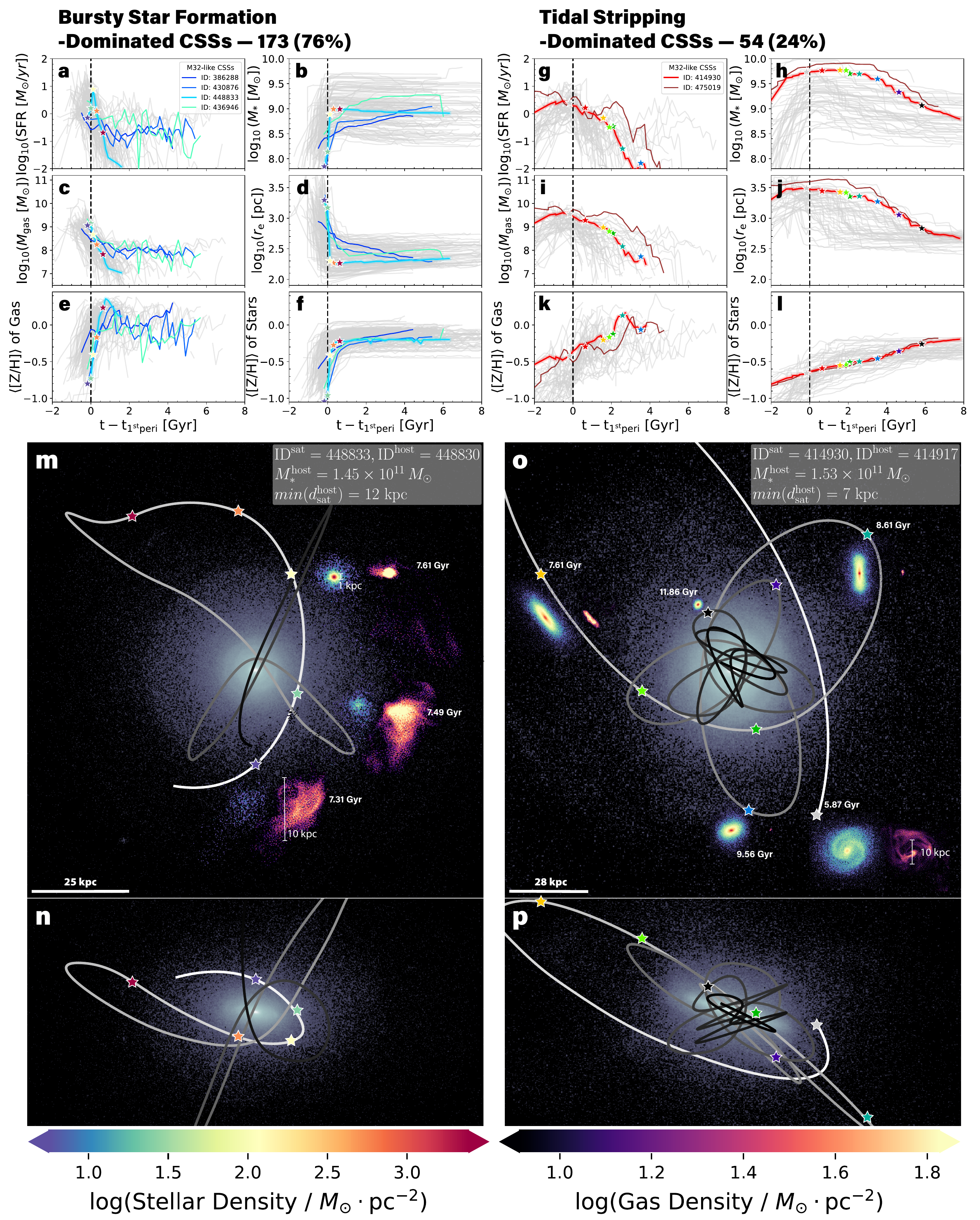}
    \caption{Top panels: the evolutionary pathways of BSF-CSSs (left, panels (a)–(f)) and TS-CSSs (right, panels (g)–(l)). Panels (a)–(f)/(g)–(l) show the evolution, centering at the time of the first pericenter ($t_{1^{\mathrm{st}}{\mathrm{peri}}}$). $M_{\mathrm{gas}}$ is the gas mass. The $\langle\text{[Z/H]}\rangle$ values of gas and stars represent the mass-weighted average metallicities of star-forming gaseous cells and star particles, respectively. Thin curves represent individual galaxies. The colored lines represent M32-like CSSs. Bottom panels: the orbits and stellar and gaseous surface density maps of examples of a BSF-CSS (left, ID 448833, panels (m)–(n)) and a TS-CSS (right, ID 414930, panels (o)–(p)). Interpolation is used to reconstruct the orbits, with a darker color indicating a later time. The colored stars mark positions in their evolution. The background shows the stellar mass surface density distribution of the host galaxy, shown in both face-on (upper) and edge-on (lower) projections at $z = 0$. The inset images, showing the stellar and gas mass surface density distribution of the CSS, are overlaid onto the background images in the face-on views. Panel (m) presents these maps at three distinct time points, and panel (o) at five. For better visibility, the stellar and gas maps are projected at a different spatial scale of 15 kpc by 15 kpc centered on the  CSS. At each time, the stellar and gas maps are horizontally offset from each other. In the upper right corner of panel (m)/(o), we list the ID for the CSS ($\mathrm{ID}^{\mathrm{sat}}$) and the host galaxy ($\mathrm{ID}^{\mathrm{host}}$), along with the stellar mass of the host galaxy ($M^{\mathrm{host}}_{*}$) and the minimum distance to the host \( min \left( d^{\mathrm{host}}_{\mathrm{sat}} \right) \).}
    \label{BSF_TS}%
\end{figure*}

In panels (a)–(f) and (g)–(l) of Figure~\ref{BSF_TS}, we present the evolution of several physical properties of CSSs, distinguishing between two formation pathways. We identify four M32-like CSSs dominated by BSF and two dominated by TS, represented by the colored lines in Figure~\ref{BSF_TS}. The orbits and morphology of two typical CSSs are shown in the bottom panels. The SFR in BSF-CSSs increases significantly during the first pericentric passage (Figure~\ref{BSF_TS}(a)). These star formation bursts generally last $\lesssim1$ Gyr, yet they lead to dramatic changes in the stellar mass content, morphology, and chemical composition of the CSSs. In comparison, TS-CSSs exhibit less bursty SFRs during their first pericentric passage (see panel (g)). Although BSF-CSSs assemble their masses later (see Figure~\ref{host}(c)), their masses grow more rapidly during their first pericentric passages than TS-CSSs (see panels (b) and (h) of Figure~\ref{BSF_TS}). In addition, BSF-CSSs retain more than half of their peak stellar masses, while TS-CSSs lose roughly 60\%–90\% of their peak stellar masses by $z = 0$. The gas mass, Mgas, and effective radius, $r_{e}$, of BSF-CSSs also decrease faster (panels (c) and (d) of Figure~\ref{BSF_TS}), indicating rapid gas consumption. As shown by the inset images in panels (m) and (n) of Figure~\ref{BSF_TS} from $t = 7.31$ Gyr (purple star) to $t = 7.61$ Gyr (yellow star), high-density gas clouds collapse, enhancing the stellar density in the central region. Instead, TS-CSSs undergo slow morphological transformations, requiring multiple pericentric passages to lose substantial mass, as illustrated by panels (o) and (p) of Figure~\ref{BSF_TS}. For BSF-CSSs, the metallicity of the gas increases by 1 dex and that of the stars by 0.8 dex, as shown by panels (e) and (f), while the gas metallicity of TS-CSSs does not increase significantly during the first pericentric passage (panel (k)). Additionally, it is clear that the increase in stellar metallicity is much slower for TS-CSSs (panel (l)).

\subsection{Highly Eccentric Orbits}\label{Orbit} 

\begin{figure}
    \centering
    \includegraphics[width=1.0\linewidth]{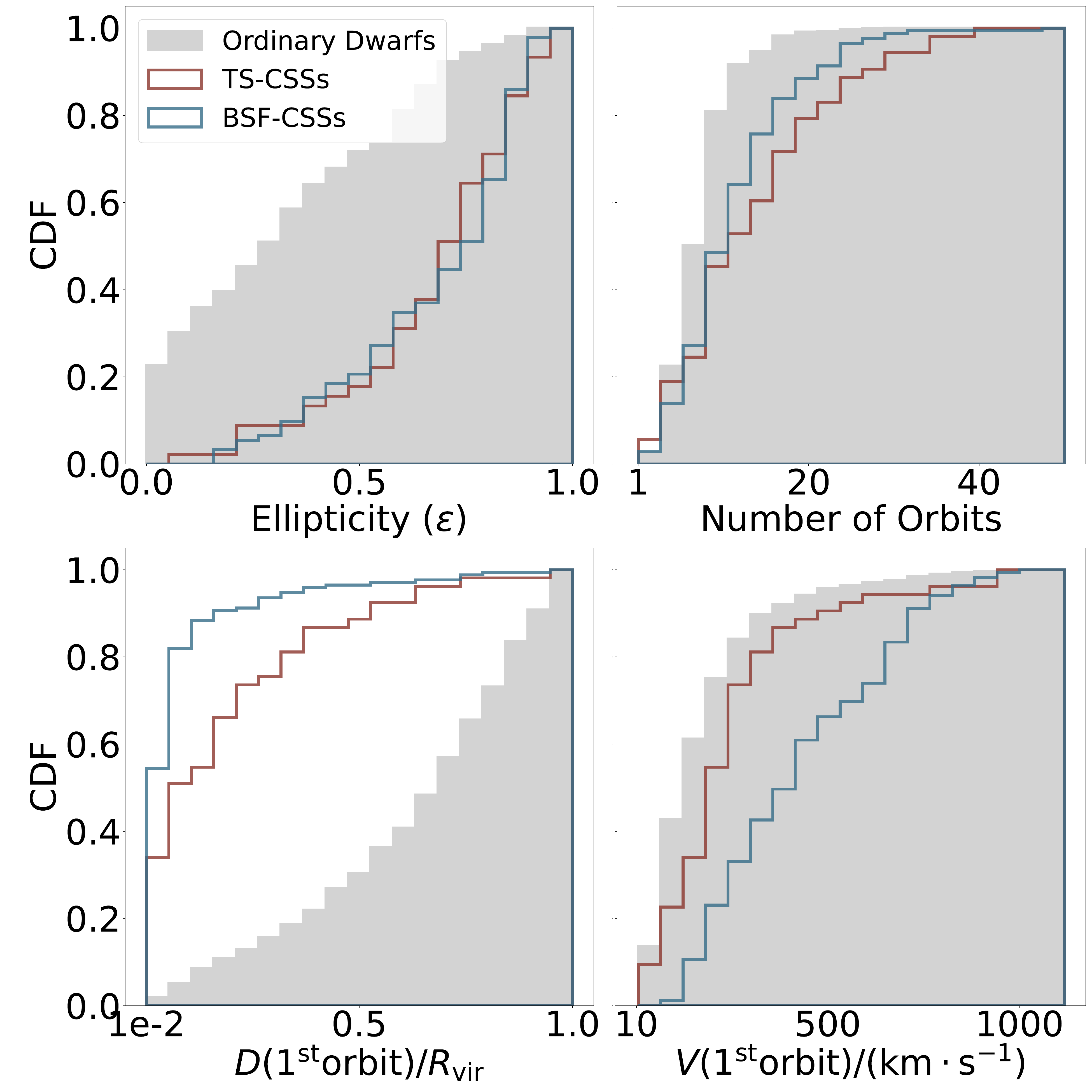}
    \caption{Orbital characteristics of CSSs selected based on our criteria in TNG50. The four panels display cumulative distribution functions for various orbital characteristics of CSSs, which include ellipticity, number of orbits, the median distance of each galaxy from its host galaxy during the first orbit ($D(1^{\mathrm{st}}\mathrm{orbit})/R_{\mathrm{vir}}$), and the median velocity during the first orbit ($V(1^{\mathrm{st}}\mathrm{orbit})$). The cumulative distribution functions for BSF-CSSs and TS-CSSs are shown in blue and red, respectively, which allows for a direct comparison against the background distribution of ordinary satellite dwarf galaxies (excluding isolated dwarfs), shown in gray. The ellipticity, denoted as \( \epsilon \), is defined as the median value of \( 1 - \frac{b}{a} \), where \( a \) and \( b \) represent the apocenter and pericenter of the orbit, respectively.}
    \label{Firstperi_orbit}%
\end{figure}

High ram pressure and BSF are expected to become important when dwarf galaxies are close to their massive hosts. This is most efficient if the dwarf galaxies are on highly eccentric orbits that bring them close to the host galaxy at high velocities \citep{Mayer2010,McGaugh2010,Du2019}. We compare the orbital characteristics of our CSSs (blue and red histograms) with those of ordinary satellite dwarf galaxies (gray, excluding isolated ones) in Figure~\ref{Firstperi_orbit}. Most CSSs follow orbits with $\epsilon > 0.5$ (the top left panel), which is determined by the median value of the ratio of the closest and farthest points of their trajectories. Moreover, CSSs complete more orbital cycles than ordinary dwarfs by counting the number of apocenters along their trajectories, shown in the top right panel. In the bottom two panels of Figure~\ref{Firstperi_orbit}, it is evident that a significant portion of CSSs undergo high-velocity close interactions with their host galaxies. The median speeds of first pericentric passage $V(1^{\mathrm{st}}\mathrm{orbit}) \simeq 220-420\ \mathrm{km/s}$ and the median distances $D(1^{\mathrm{st}}{\mathrm{orbit}})/R_{\mathrm{vir}} \simeq 0.05-0.08$ during the first orbit. CSSs thus experience a substantially higher ram pressure than ordinary dwarfs.

\subsection{Metal Enrichment Caused by Ram Pressure Confinement}\label{CSS-BSF}

\begin{figure*}
    \centering
    \includegraphics[width=1\linewidth]{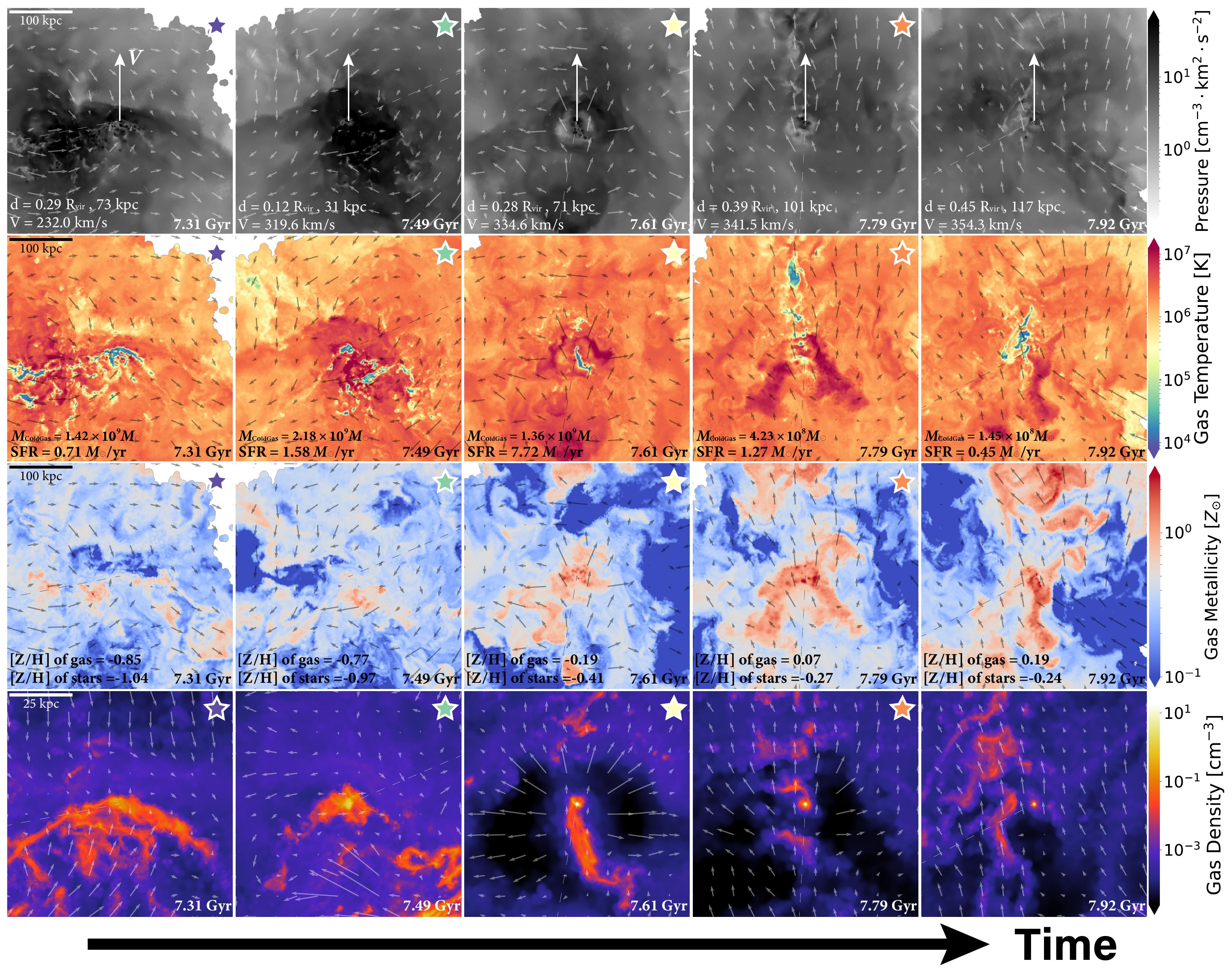}
    \caption{Illustrative evolution of the gas for a BSF-CSS during the first pericentric passage. The ID of the CSS at $z=0$ is 448833. The CSS is centrally positioned  within each panel, with time annotated in the lower right corner. Each panel (except for those in the bottom row) covers an area of $400\times400\ \text{kpc}^2$, projected onto a plane perpendicular to the bulk velocity vector direction of the CSS and rendered at a resolution of $800\times800$ pixels. Dashed lines connect the position of the CSS to its host, while arrows denote the velocity direction of the CSS relative to the host. Quiver plots are overlaid on each panel, illustrating the gas velocity field around the CSS. The top row shows the pressure field, measured in dynes per square centimeter, with distance and velocity indicated in the lower left corner of each panel. The second row presents the gas temperature, with the mass of cold gas (identified by a star formation rate of the cell $> 0$) and the SFR shown in the lower left corner of each panel. The third row shows the evolution of gas metallicity, with the $\langle \text{[Z/H]} \rangle$ value of stars and $\langle \text{[Z/H]} \rangle$ value of gas specified in the lower left corner of each panel. The bottom row zooms in on the gas density distribution.}
    \label{BSF_gas_evo}%
\end{figure*}

\begin{figure*}
    \centering
    \includegraphics[width=1\linewidth]{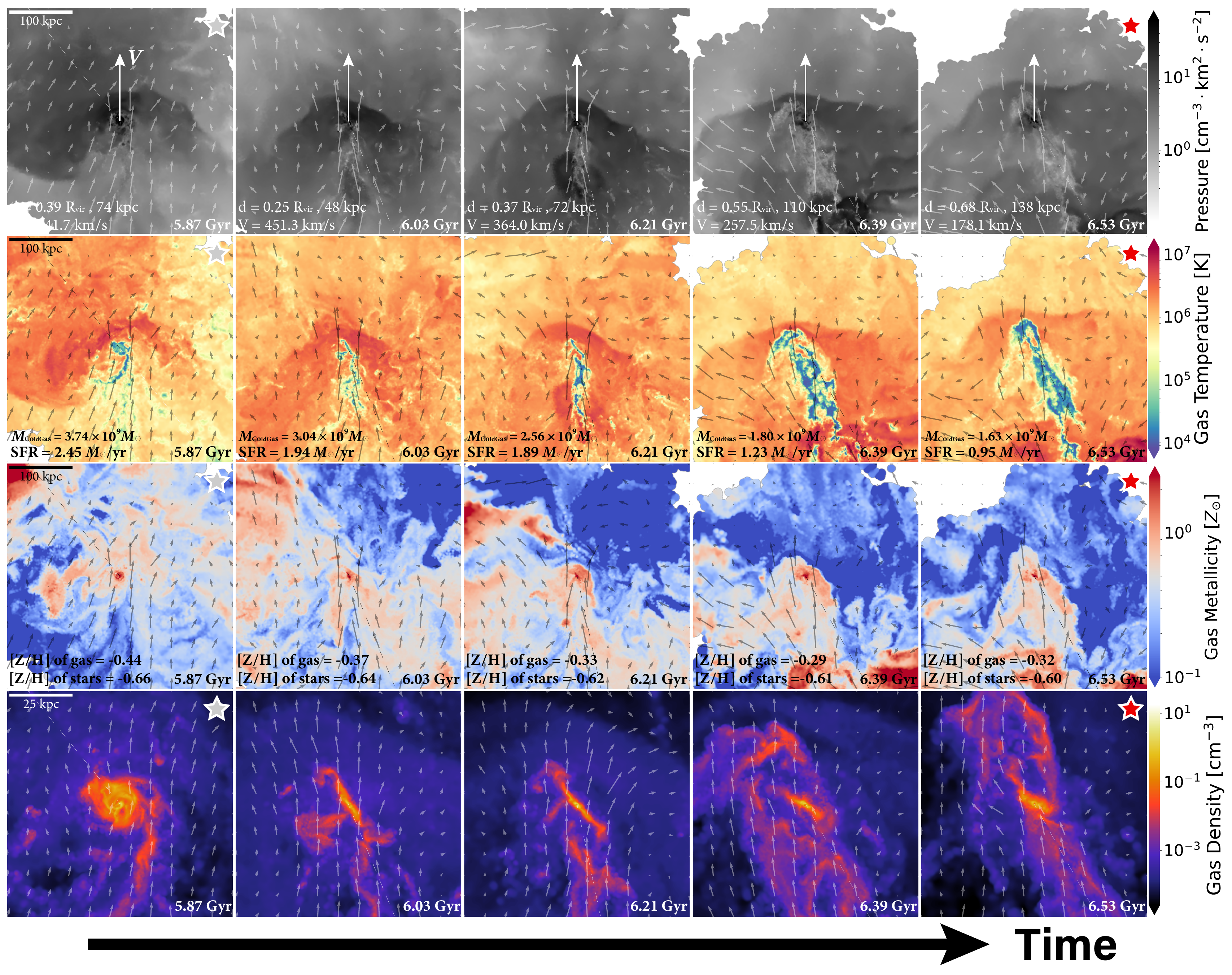}
    \caption{Illustrative evolution of the gas for a TS-CSS during the first pericentric passage. The ID of the CSS at $z = 0$ is 414930. This figure uses the same conventions as those established in Figure~\ref{BSF_gas_evo}.}
    \label{TS_gas_evo}%
\end{figure*}

\begin{figure}
    \centering
    \includegraphics[width=1\linewidth]{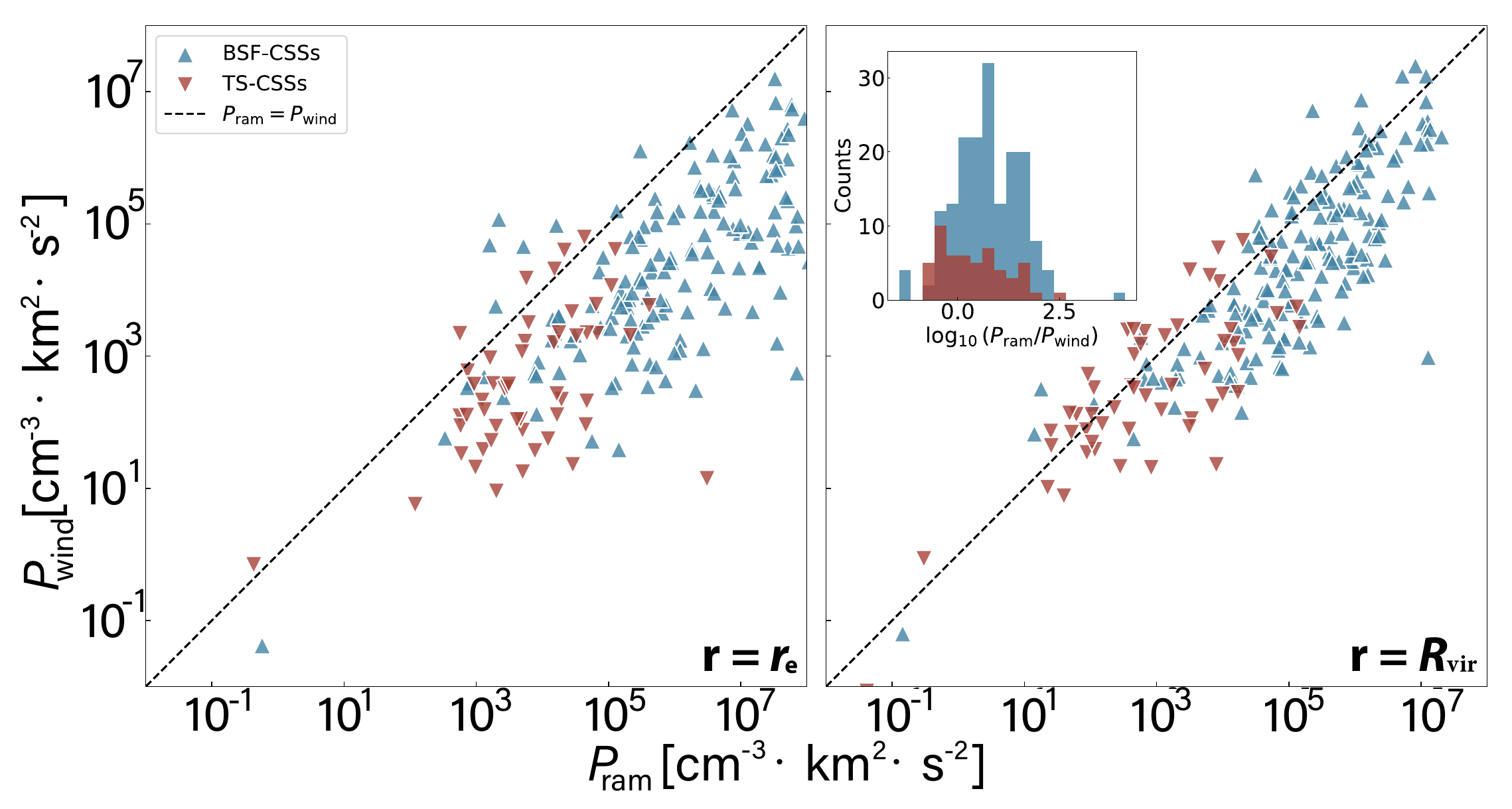}
    \caption{$P_{\text{ram}}$ vs. $P_{\text{wind}}$ of each CSS at two radii, $r_{e}$ (left) and $R_{\mathrm{vir}}$ (right), during its first pericentric passage. The figure compares the ram pressure, $P_{\text{ram}}$, experienced by CSSs as they move through the surrounding medium (defined by Equation~\ref{P_ram}), with the pressure exerted by galactic winds, $P_{\text{wind}}$, as defined by Equation~\ref{P_wind}. The data is shown with blue for BSF-CSSs and red for TS-CSSs. The inset image shows the distribution of log $(P_{\text{ram}}/P_{\text{wind}})$ at $R_{\mathrm{vir}}$.}
    \label{Pressure_comparison}%
\end{figure}

We use two M32-like CSSs, one a BSF-CSS and the other a TS-CSS, to illustrate the chemodynamic evolution during their first pericentric passage. Figures~\ref{BSF_gas_evo} and Figures~\ref{TS_gas_evo} show the metal enrichment resulting from ram pressure confinement, evident in  the gaseous pressure (top row)\footnote{Noted that in Figures~\ref{BSF_gas_evo}, ~\ref{TS_gas_evo}, and ~\ref{Pressure_comparison}, we represent pressure in units of $\mathrm{cm}^{-3} \cdot \mathrm{km}^{2} \cdot \mathrm{s}^{-2}$. This unit is chosen based on our definition of gas density as the hydrogen number density ($n_{\mathrm{H}}$) in $\mathrm{cm}^{-3}$ and the velocity squared ($v^2$) in $\mathrm{km}^{2} \cdot \mathrm{s}^{-2}$.}, temperature (second row), metallicity (third row), and “zoom-in” density (bottom row) distributions during the first pericentric passage. We orient the galaxies such that their velocity (the white arrows) is along the y-axis. Their morphology is characteristic of ``jellyfish galaxies," which display long trailing gas tails \citep[e.g.,][]{Rohr2023,Zinger2024}. In each case, a clear bow shock \citep[e.g.,][]{Binney1987,Yun2019} appears ahead of the galaxy owing to its supersonic velocity. The high ram pressure, resulting from the CSSs’ rapid motion near the halo center, contributes to enhancing SFR by compressing the gas \citep[e.g.,][]{Kronberger2008,Ramatsoku2020,Vulcani2020,Goller2023}. The gaseous quiver plots illustrate that ram pressure compresses gas on the leading edge of the CSS (relative to its direction of motion), as indicated by inward flow vectors in the gas velocity field on this side, as seen in the first two columns of Figures 8 and 9. Gas outflows are preferentially suppressed in the direction opposite to the ram pressure force. This results in a highly anisotropic velocity field, with little to no outward flow along the direction of the pressure. The confined gas escapes more easily in directions perpendicular to the CSS motion, as indicated by the higher-velocity components along the x-axis (third column of Figure~\ref{BSF_gas_evo} and fourth column of Figure~\ref{TS_gas_evo}). On the trailing side, the gas forms wakelike structures as it flows around the CSS, forming confined outflows and then becoming a tail of stripped gas. As a result, some of the metal-enriched hot gas from the BSF is confined to the vicinity of the CSS, where it can be recycled into the next generations of stars, enabling them to reach high metallicity rapidly. The confined gas further fuels star formation in the central regions, thereby helping to rapidly transform a metalpoor but gas-rich dwarf galaxy into a metal-rich CSS, as seen in panel (f) of Figure~\ref{BSF_TS}.

We compare the strength of ram pressure and outflow winds around CSSs in Figure~\ref{Pressure_comparison} to determine whether ram pressure confinement can sufficiently suppress supernova-driven outflows. This assessment is based on a comparison of ram pressure ($P_{\text{ram}}$) and feedback wind pressure ($P_{\text{wind}}$) at two radii, $r_{e}$ and $R_{\mathrm{vir}}$ of each CSS, during the first pericentric passage. As outlined in \citet{Gunn&Gott1972}, the ram pressure $P_{\text{ram}}$, is given by:
\begin{equation} \label{P_ram}
P_{\text{ram}} = \rho_{\text{ambient}} \cdot v_{\text{bulk}}^2
\end{equation}
where $v_{\text{bulk}}$ is the relative velocity of a galaxy to its ambient gas. Meanwhile, $\rho_{\text{ambient}}$ is approximated by averaging the density measurements taken between $R_{\text{vir}}$ and $2R_{\text{vir}}$ encircling each CSS. The pressure from galactic winds, $P_{\text{wind}}$, is estimated by 
\begin{equation} \label{P_wind}
P_{\text{wind}} = \frac{\dot{M}_{\text{wind}} v_{\text{wind}}}{4 \pi R_{\text{M}}^2} = \overline\rho_{\text{wind}}\cdot v_{\text{wind}}^2
\end{equation}
where $\dot{M}_{\text{wind}}$ is the mass outflow rate. We evaluate the effect of ram pressure around the virial radius of CSSs, i.e., $R_{M} = R_{\mathrm{vir}}$. 
Note that only gas reaching $R_{\mathrm{vir}}$ with $v > v_{\mathrm{esc}}$ can escape the potential well of CSSs. The mass outflow rate is calculated as $\dot{M}_{\text{wind}} = 4 \pi R_{M}^2\  \overline\rho_{\text{wind}} \cdot v_{\text{wind}}$, where $\overline\rho_{\text{wind}}$ represents the mean density of the expelled gases within a shell width $0.01R_{\mathrm{vir}}$ at $R_{\mathrm{vir}}$. $v_{\text{wind}}$ is estimated by the mean velocity of gas particles that exceed the local escape velocity. Figure~\ref{Pressure_comparison} demonstrates that $P_{\text{ram}}$ is generally much greater than $P_{\text{wind}}$ across a wide spatial scale, indicating that ram pressure is sufficient to confine outflows when CSSs are in close proximity during their encounters. The plausible impact of ram pressure in stimulating star formation and enhancing metallicity has also been discussed in several early studies. \citet{Babul1992} speculated that the location of a galaxy within a cluster could determine whether it evolves into a nucleated \textit{dE}. \citet{Murakami&Babul1999} explored this further, examining the effects of supernova winds and confinement through hydrodynamical simulations. However, the detailed results from these studies were inevitably constrained by the limitations of using simplified spherical models and the basic 2D hydro-code available at the time. As early as \citet{Schulz&Struck2001}, the idea of ram pressure-enhanced star formation was discussed. \citet{Du2019} further demonstrated that in their advanced smooth-particle hydrodynamical simulation, through tracing gas particles, a large proportion of metal-rich gas ejected by supernova feedback can return to the dwarf galaxy under a high ram-pressure environment.

As shown in Figure~\ref{TS_gas_evo}, the ram pressure confinement may play a relatively minor role in the metal enrichment of TS-CSSs. The progenitors of TS-CSSs have formed most of their stars already by the time of the first pericentric passage. These progenitors form more massive and stable galactic structures
earlier in their evolution, and therefore, at the time when ram pressure begins to act, a substantial population of older, metal-poor stars is already in place. Consequently, the enrichment in TS-CSSs
proceeds more slowly, as the relative impact of ram pressure confinement is less pronounced in systems that already host significant stellar populations and exhibit more gradual star formation histories. In this dense environment, the strangulation effect \citep{Peng2015,Pasquali2012} and high thermal pressure \citep{Petropoulou2012} near a massive galaxy may also contribute to metal enrichment, but less significantly than ram pressure.

\subsection{Stellar Population Diagnostics of the Two Formation Channels} \label{Fingerprint}

\begin{figure}
    \centering
    \includegraphics[width=1.0\linewidth]{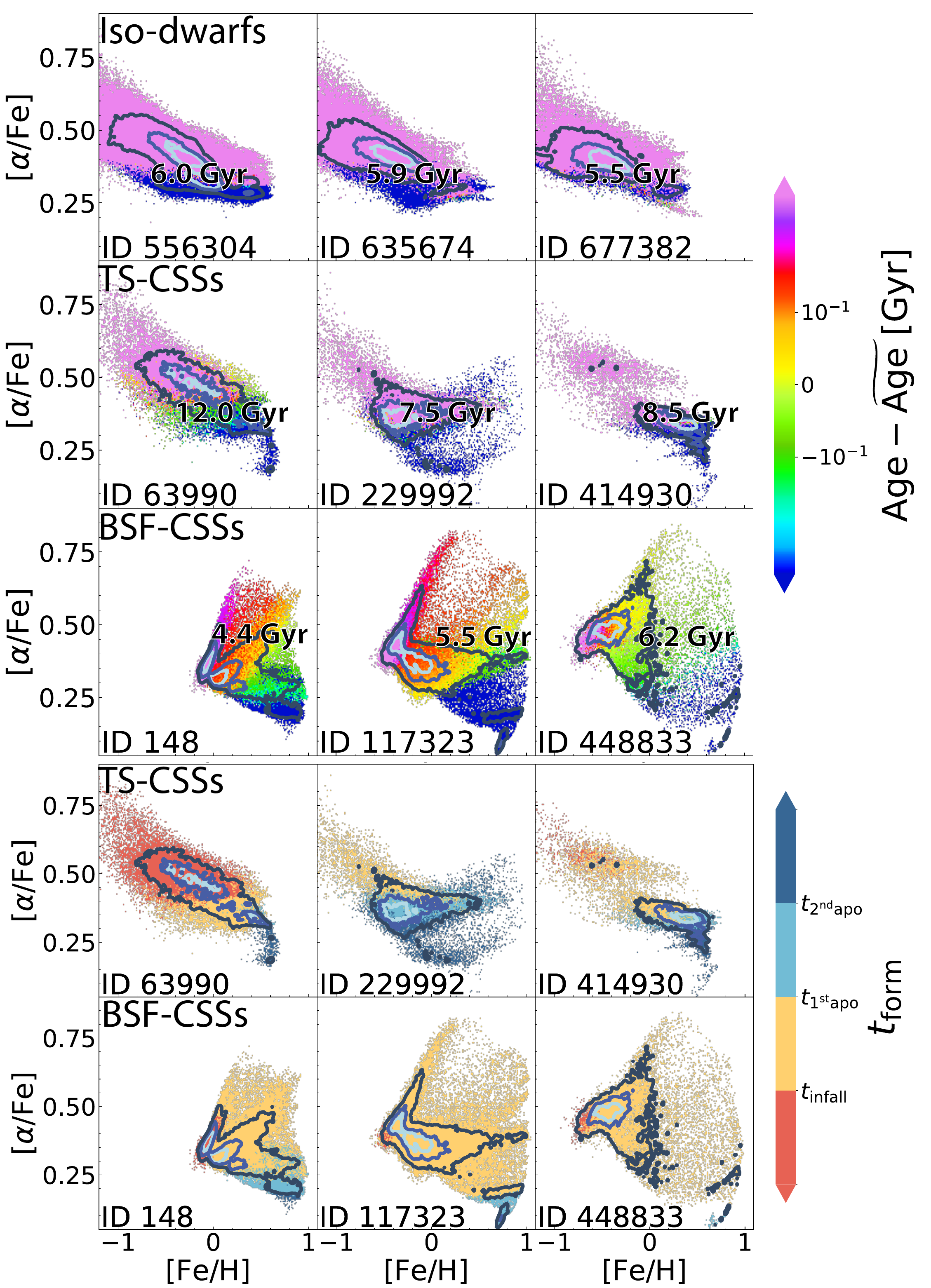}
    \caption{$\mathbf{[\alpha/\mathrm{Fe}]}$-$\mathrm{[Fe/H]}$ diagram of stellar populations of CSSs at $z=0$ formed by different mechanisms. $\alpha$ species are \ce{^{12}C}, \ce{^{16}O}, \ce{^{20}Ne}, \ce{^{24}Mg}, \ce{^{28}Si}. The contours from light to dark blue represent the maximum density's 20, 40, and 80 percentiles, respectively. In the upper nine panels, from top to bottom, we present cases of isolated central dwarfs (Iso-dwarfs), TS-CSSs, and BSF-CSSs, with stars color-coded based on the difference between their age and the median age ($\widetilde{\mathrm{Age}}$) of the stellar populations in their respective galaxies. The $\widetilde{\mathrm{Age}}$ for each example is indicated within its corresponding panel. In the lower six panels, stars are color-coded based on their respective formation periods to illustrate the timing of the formation of distinct stellar populations: red (stars formed before infall), orange (during the first orbit), light blue (during the second orbit), and navy (beyond the second orbit). These distributions exhibit significant differences, with BSF-CSSs exhibiting distinct finger features extending towards high $\mathbf{[\alpha/\mathrm{Fe}]}$ and $\mathrm{[Fe/H]}$.}
    \label{alphaFe_FeH}%
\end{figure}

Stellar populations can be used to differentiate between the two formation channels. The top panels of Figure~\ref{alphaFe_FeH} show three example chemical maps of isolated dwarf galaxies. In these galaxies, old stars (pink color) have low metallicity and high $\alpha$-abundance, while younger stars (dark blue color) have higher metallicity but lower $\alpha$-abundance. This trend arises from the less intense star formation at late times. The chemical composition of BSF-CSSs, however, exhibits the finger features predicted by \citet{Du2019}, as can be seen in the bottom row of the upper panels of Figure~\ref{alphaFe_FeH}. The generation of these structures is a result of the bursty star formation and rapid enrichment during the first pericentric passage (yellow regions in the lower panels) of CSSs.

The finger features are common in BSF-CSSs (135/173) while 6/54 of TS-CSSs also have similar finger features (see one example in the middle panel of the second row of Figure~\ref{alphaFe_FeH}), implying that ram pressure confinement also influences the formation of new stars in the central regions of some TS-CSSs. Although these finger features have been found in very different simulated galaxies -- the current work, and \citet{Du2019} -- they have not yet been observed in galaxies such as M32 due to current observational difficulties. Detailed $[\alpha/\mathrm{Fe}]$ ratio measurements are only feasible in the Milky Way and a few nearby galaxies, where spectral data of individual stars can be obtained with sufficient quality. However, exploring similar features in more distant galaxies such as M32 awaits the advanced capabilities of instruments like ELT/\textit{JWST}. A recent analytical study by \citet{Ting2024} employed Gaussian mixture models to search for multimodal, discontinuous tracks in the [Mg/Fe]-[Fe/H] plane using an unsupervised approach. They also demonstrates that bursty star formation in dwarf galaxies produces a distinct, discontinuous chemical track, clearly distinguishable from the smoother track seen in dwarf galaxies with continuous star formation. We predict that objects affected by ram pressure confinement will exhibit finger features, with stars concentrated in regions of intense star formation. During their formation, metallicity and $[\alpha/\mathrm{Fe}]$ rise rapidly due to Type II supernovae, but decrease when Type Ia supernovae begin after a few hundred Myr. This would create an age sequence in the finger features, with stars getting younger from the base to the tip, as shown in the bottom row of the upper panels of Figure~\ref{alphaFe_FeH}.

Another distinctive difference between the two types of CSSs is the scarcity of old stars in BSF-CSSs, due to their relatively late formation, as shown in Figure~\ref{Age_Fe_alpha}. Conversely, TS-CSSs exhibit a significantly reduced population of young stars compared to isolated dwarfs, a consequence of ram-pressure stripping (RPS), as illustrated in panel (c) of Figure~\ref{host}.

\section{Discussions} \label{sec:discussion}

\subsection{Origin of Stellar Halo and Giant Stellar Streams in a M31-M32 Analog System}

\begin{figure*}
    \centering
    \includegraphics[width=1\linewidth]{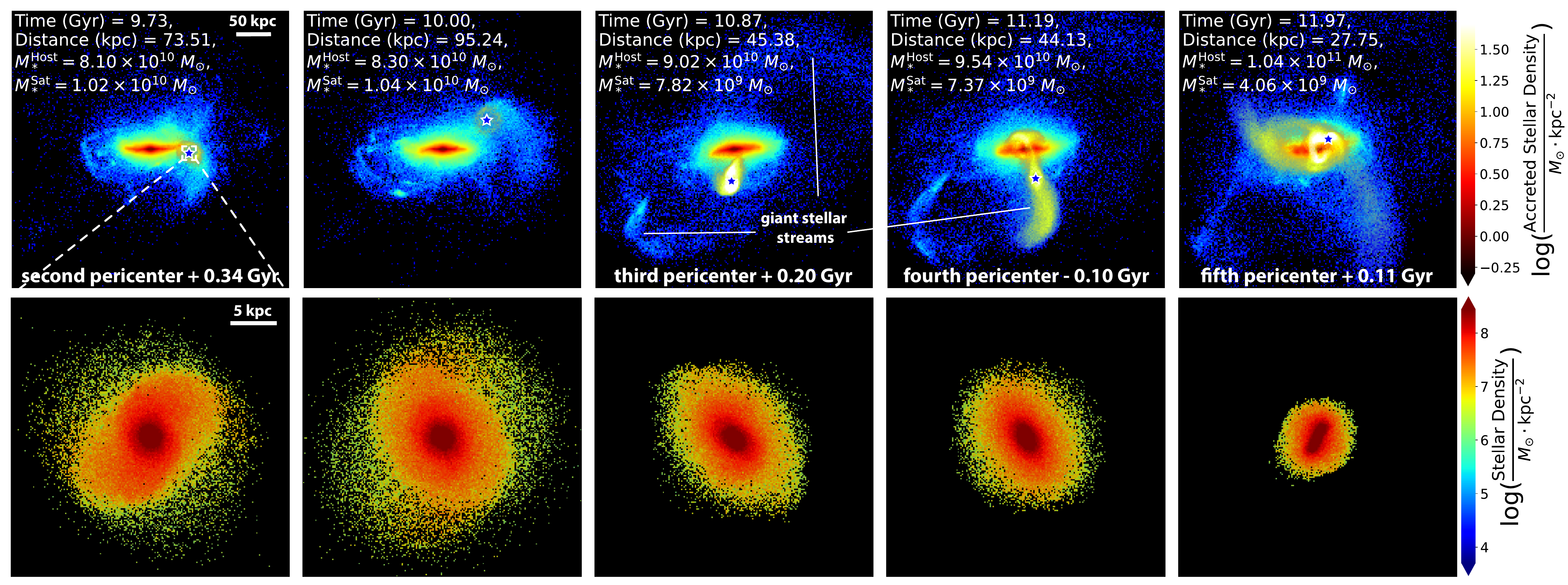}
    \caption{Formation of the stellar halo in a TS-dominated M31-M32 analog system. Here we select the time when the CSS is interacting with the central host, resembling the M31-M32 system. The first row shows the stellar halo formed through close interactions. The stellar halo is comprised of material from both the central galaxy (depicted in blue colors) and from the progenitor of the CSS (depicted using yellow and red colors). The second row's detailed images display the stellar surface density of the CSS during these interactions.}
    \label{Host_475016_Sat_475019}%
\end{figure*}

In TNG50, TS-CSSs undergo slow morphological transformations due to multiple events that lead to significant mass loss. This process is illustrated in panels (o) and (p) of Figure~\ref{BSF_TS} and further supported by Figure~\ref{Firstperi_orbit}. This contrasts with previous models, such as \citet{DSouza2018}, who suggest that M32 formed rapidly due to a single tidal interaction approximately 2 Gyr ago. They further argued that the giant stellar stream is originated from the tidal interaction between M31 and M32 .

We identified a fairly realistic M31-M32 analog system (ID 475016-ID 475019) shown in Figure~\ref{Host_475016_Sat_475019}, where at low redshifts, tidally induced giant stellar streams form, similar to observed structures \citep{Ibata2001, Ferguson2002, Zucker2004}. These streams, which include material from the host galaxy (shown in blue and green) and the tidally stripped debris of the CSS progenitor (shown in yellow and red), phase mix over time. However, the massive CSS progenitor ($M_{*} \simeq 10^{10} M_{\odot}$), generates a giant stellar stream from the host galaxy, resulting in a stellar halo of too high a mass, with the outer halo ($R \gtrsim 30\ \text{kpc}$) having a mass of $2.98 \times 10^{10} M_{\odot}$ at $z=0$, whereas observations of M31 suggest a total stellar halo mass of $\simeq 2.1^{+1.7}_{-0.4} \times 10^{9} M_{\odot}$ \citep{Williams2012}. This discrepancy suggests that the TS scenario is unlikely to be the cause of M32. We conclude that the BSF scenario offers a more likely explanation for the observed relatively low-mass stellar halo of M31.

\subsection{Paucity of Dark Matter}

It has been noted that there are CSSs that lack dark matter in different cosmological simulations \citep{Jang2024}. They are associated with massive hosts and are relatively metal-rich. The `missing dark matter' in simulations of CSSs suggests that either these systems have formed galaxies with unusually high efficiency or their dark matter content is significantly lower than expected from $\Lambda$CDM. Notably, \citet{Lora2024} report that the process seen in ``jellyfish" galaxies in TNG—where stripped gas in the tails compresses and triggers star formation—could also apply to CSSs. These regions can evolve into self-gravitating dwarf galaxies. Our results suggest that BSF-CSSs are formed from much less massive progenitor galaxies ($\sim 10^{7-8} M_{\odot}$), possibly with smaller dark matter halos than expected, due to enhanced star formation efficiency resulting from close interactions. Furthermore, multiple close interactions with their massive host galaxies result in substantial stripping of dark matter from the CSSs \citep{Smith2016b,Hammer2020,Keim2022}. Moreover, strong stellar feedback within dwarf galaxies induces significant fluctuations in the gravitational potential by expelling gas, potentially affecting the orbits of dark matter particles and reducing its density \citep{Navarro1996,Gelato1999,Read2005,Pontzen2012,Sales2022}. In addition, as mentioned earlier in Section~\ref{classification}, by tracing particles, we found that a small portion of BSF-CSS progenitors originated from material stripped from satellite galaxies orbiting central galaxies. These fragments continued their orbits around the central galaxies over time.

\subsection{Physical versus Numerical Origins of UCDs and cEs}

\citet{Nelson2019-1} details the classification of subhalos in TNG simulations, distinguishing some subhalos as non-cosmological `clumps' formed from baryonic processes and flagging them for careful analysis, due to their typical characteristics of being low-mass, baryon-dominated, and centrally located in host galaxies at $z < 1$. \citet{Boecker2023} further showed that these so-called `clumps' in TNG50 migrate towards the galaxy center, thereby enhancing the central concentration of stars and gas during mergers or close encounters. In our study, the 147 CSSs not tracked by the default \texttt{SubLink} algorithm correspond to these flagged subhalos. We have extended the evolutionary history of these objects through star particle matching, capturing their earlier formation periods, and argue for their physical origins and relevance as real CSSs counterparts. Contrary to expectations for mere `clumps' derived from the disks of massive galaxies, which would typically inherit high metallicities at birth, these objects instead exhibit significant subsequent enrichment, as shown in Figure~\ref{classification}. Furthermore, if these systems were merely numerical clumps of gas and stars, we would expect them to lack the well-defined evolutionary histories observed in the CSSs. In contrast, the CSSs in our sample exhibit coherent baryonic mass assembly histories and have well-defined orbits and infall trajectories, further supporting their interpretation as physical systems.

It is important to distinguish our sample from \citet{Goller2023}, who focused on typical ``jellyfish galaxies" and found no population-wide SFR enhancement by ram pressure in TNG50. Unlike their sample, our galaxies are smaller and more gas-rich, making them more sensitive to ram pressure. The highly eccentric orbits of the BSF-CSSs in our study (Figure~\ref{Firstperi_orbit}) result in shorter pericentric passage times, leading to stronger and more concentrated ram pressure compression. This enhanced compression efficiently compresses gas and significantly boosts star formation.

Our findings align to some extent with previous studies \citep[e.g.,][]{Lee2020,Choi2022,Zhu2024}, who report short-lived and mild SFR enhancements due to ram pressure in controlled simulations. However, the galaxies in those studies are typically more massive, similar to the progenitors of TS-CSSs during their first pericentric passage. As a result, the SFR enhancements in those studies are weaker compared to the stronger enhancements in our BSF-CSSs sample.

The existence of CSSs of even lower mass ($\sim 10^{7} M_{\odot}$) in the \texttt{NewHorizon} simulation further corroborates our findings, despite its different numerical methodologies and treatments of stellar feedback. Given the lack of a clear physical boundary distinguishing UCDs from cEs, it is reasonable to hypothesize a similar evolutionary pathway for both categories. \citet{Jang2024} identified a subset of CSSs with a stellar mass in the range of $M_{*} \sim 10^{6}-10^{9}\ M_{\odot}$ in the high-resolution \texttt{NewHorizon} simulation. The (highest) mass resolutions of the dark matter and stellar particle are $1.2 \times 10^{6}\ M_{\odot}$ and $1.3 \times 10^{4}\ M_{\odot}$ for each, respectively. Also, the corresponding maximum spatial resolution for the mesh structure is 34 pc at $z = 0$. This led to the discovery of more UCDs in their CSSs sample. They characterized some of them as `intrinsic-associated', highlighting their metal abundance and scarcity of dark matter. They suggest that these obejcts appear to have emerged from a rapid starburst episode. Additionally, the ratio of `intrinsic' to `stripped' CSSs in their analysis 70\% to 30\% matches well with our findings in TNG, with a 76\% to 24\% division. This agreement of independent studies corroborates the universality of the processes shaping the evolution of both UCDs and cEs.

\subsection{Numerical Resolution and TNG Model Caveats}

\citet{Pillepich2019} argues that the sizes of galaxies with $M_{*} \gtrsim 10^{8} M_{\odot}$ are reasonably well resolved in TNG50. However, its limited resolution still poses challenges when studying CSSs, potentially affecting these systems. TNG50 uses a Plummer-equivalent gravitational softening length of 0.29 kpc at $z=0$. This gives a lower limit to the galaxy sizes that can be well resolved, as indicated by the horizontal dashed line in the left panel of Figure~\ref{mass-size}, which is greater than the sizes of some of the smallest CSSs in the simulation. Although the gravitational force becomes non-Newtonian at this scale, it does not significantly affect the chemical evolution of each CSS. Given these constraints, our analysis focuses on the overarching, statistically validated characteristics of CSSs. This approach ensures that our conclusions about their formation pathways—reflecting the true physical processes—are robust despite the TNG50’s finite resolution.

The simplified interstellar medium (ISM) model employed in the TNG simulations motivates additional caution \citep{Springel2003}. It assumes star formation occurs over a characteristic timescale tied to local dynamical processes such as gas flow, collisions, and compression. Thus star formation rates can vary significantly with environmental conditions. Shorter dynamical times indicate faster gas collapse, potentially leading to rapid star formation, especially in high-density areas. The stellar feedback is implemented through the generation of decoupled kinetic winds, which are proportional to the prompt energy released by Type II supernova explosions \citep{Pillepich2018a}. In this regime, the feedback-driven wind exits the ISM of the galaxy without direct interaction, potentially facilitating the artificial formation of compact galaxies. For instance, a more resolved feedback model \citep[e.g.,][]{Smith2021}, could disrupt the star-forming ISM of CSSs more effectively or earlier, thereby suppressing their growth. Future simulations with more detailed star formation and stellar feedback models will further illuminate the formation pathways of CSSs.

\section{Summary} \label{sec:summary}

In this paper, we investigate the effects of ram pressure and tidal stripping on the formation of metal-rich CSSs in the TNG50 simulation. By tracing back the evolutionary histories of all 227 CSSs, we identify and reassess the prevalence of the two primary mechanisms for CSSs formation—bursty star formation (76\%) and tidal stripping (24\%)—within a cosmological context. From gas phase metallicity to stellar metallicity, we showed that the effect of ram pressure confinement on the metallicity of CSSs is significant. The two formation scenarios can be summarized as follows:
\begin{enumerate}
    \item[(1)] \textit{BSF-CSSs:} Gas-rich dwarf galaxies, falling into the vicinity of a massive host during the first pericentric passage, undergo the combined effects of ram pressure compression and tidal stirring, which trigger bursts of star formation. This leads to a rapid star formation which transforms gas-rich dwarfs into metal-rich and compact stellar systems. Though a significant portion of the gas is stripped, the confinement effect caused by strong ram pressure due to high-speed close flybys significantly suppresses metal-rich feedback-driven outflows, enabling the infalling dwarf galaxy to retain a higher metal content than galaxies of comparable masses.
    \item[(2)] \textit{TS-CSSs:} Typical star-forming galaxies, upon falling into a galaxy cluster, gradually quench due to the effects of ram pressure stripping, with ram pressure compression also stimulating star formation to a certain extent. After quenching, they undergo frequent and strong tidal interactions that gradually strip away the diffuse outskirts, which are primarily composed of old, metal-poor, and $\alpha$-rich stars. Bursty star formation and ram pressure confinement are also important but play a relatively minor role. As a result, they become less massive, more metal-rich, but less $\alpha$-enhanced. 
\end{enumerate}
Here we have demonstrated that while tidal stripping contributes to the creation of metal-rich CSSs, it is not the primary mechanism in a cosmological setting. Rather, the compression and confinement effects resulting from ram pressure are crucial. The swift increase in metallicity observed in CSSs is a consequence of the inhibition of outflows and bursty star formation in satellite galaxies orbiting near a massive host galaxy. This situation has become more prevalent since redshift $z\sim 1$, resulting in the emergence of a younger CSS population with even higher metallicities. The bursty star formations generate a small scatter in age as well as testable finger features on the $[\alpha/\text{Fe}]-\text{[Fe/H]}$ diagram.

Our research has important implications for understanding the formation and evolution of compact stellar systems when falling into dense environments. It suggests that the progenitors of CSSs are generally gas-rich dwarf galaxies or even high-velocity clouds. We acknowledge the substantial uncertainty inherent in hydro-simulations, where the effectiveness of supernova feedback and subgrid-physics can affect the outcomes. It is essential to confirm our theoretical predictions through observational measurements.

%\section*{Acknowledgements}
\begin{acknowledgments}
We thank the anonymous referee for their helpful comments which improved the quality of this manuscript. Y.B. thanks Volker Springel for constructive comments and suggestions. Y.B. also thanks Guangwen Chen and Hong-xin Zhang for supplying the literature CSSs data. This study acknowledges the support by China Manned Space Program through its Space Application System, the Natural Science Foundation of Xiamen China (No. 3502Z202372006), the Fundamental Research Funds for the Central Universities (No. 20720230015), and the Science Fund for Creative Research Groups of the National Science Foundation (NSFC) of China (No. 12221003). L.C.H. is supported by the National Science Foundation of China (No. 11721303, 11991052, 12011540375, 12233001), the National Key R\&D Program of China (2022YFF0503401), and the China Manned Space Project (CMS-CSST-2021-A04, CMS-CSST-2021-A06). R.C. is supported in part by the National Key Research and Development Program of China. C.G. acknowledges support from the National Natural Science Foundation of China (No. 12373007, 12422302). H.L. is supported by the National Key R\&D Program of China No. 2023YFB3002502, the National Natural Science Foundation of China under No. 12373006. This study utilizes the TNG50 simulation, one of the flagship runs of the IllustrisTNG project, run on the HazelHen Cray XC40 system at the High Performance Computing Center Stuttgart as part of project GCS-ILLU of the Gauss centers for Supercomputing (GCS). The data analysis was conducted using \texttt{yt} \citep[\href{https://yt-project.org}{https://yt-project.org}, ][]{yt}, \texttt{Numpy} \citep{Harris2020}, \texttt{Scipy} \citep{Virtanen2020}, \texttt{Matplotlib} \citep{Hunter2007}, \texttt{Seaborn} \citep{Waskom2021}, \texttt{Pynbody} \citep{Pontzen2013}, \texttt{Pandas} \citep{McKinney2011} and \texttt{h5py} \citep{Collette2021}. This work was also strongly supported by the Computing Center in Xi'an.
\end{acknowledgments}

% \appendix

\bibliography{CSSs}{}
\bibliographystyle{aasjournal}

%% This command is needed to show the entire author+affiliation list when
%% the collaboration and author truncation commands are used.  It has to
%% go at the end of the manuscript.
%\allauthors

%% Include this line if you are using the \added, \replaced, \deleted
%% commands to see a summary list of all changes at the end of the article.
%\listofchanges

\end{document}